\documentclass[longauth]{aa} 

\usepackage{booktabs}
\usepackage{graphicx}
\usepackage{multirow}
\usepackage{natbib} 
\usepackage{txfonts}
\begin{document}

   \title{The CARMENES search for exoplanets around M dwarfs:}
   \subtitle{Convective shift and starspot constraints from chromatic radial velocities}

   \author{D.~Baroch\inst{1,2} \and J.~C.~Morales\inst{1,2} \and I.~Ribas\inst{1,2} \and E.~Herrero\inst{1,2} \and A.~Rosich\inst{1,2} \and M.~Perger\inst{1,2} \and G.~Anglada-Escud\'e\inst{1,2} \and A.~Reiners\inst{3} \and J.~A.~Caballero\inst{4} \and A.~Quirrenbach\inst{5} \and P.~J.~Amado\inst{6} \and S.~V.~Jeffers\inst{3} \and C.~Cifuentes\inst{4} \and V.~M.~Passegger\inst{7,8} \and A.~Schweitzer\inst{7} \and  M.~Lafarga\inst{1,2}  \and F.~F.~Bauer\inst{6} \and V.~J.~S.~B\'ejar\inst{9,10}  \and J.~Colom\'e\inst{1,2} \and M.~Cort\'es-Contreras\inst{4,11} \and S.~Dreizler\inst{3}  \and D.~Galad\'i-Enr\'iquez\inst{12} \and A.~P.~Hatzes\inst{13} \and Th.~Henning\inst{14} \and A.~Kaminski\inst{5} \and M.~K\"urster\inst{14}  \and D.~Montes\inst{15}  \and
   C.~Rodr\'iguez-L\'opez\inst{6} \and
   M.~Zechmeister\inst{3}
          }
   \authorrunning{D. Baroch et al.}
   \titlerunning{Stellar activity constraints from chromatic radial velocities}
   \institute{Institut de Ci\`encies de l'Espai (ICE, CSIC),
              Campus UAB, c/Can Magrans s/n, E-08193 Bellaterra, Barcelona, Spain\\
              \email{baroch@ice.cat}
         \and
              Institut d'Estudis Espacials de Catalunya (IEEC),
              c/ Gran Capit\`a 2-4, E-08034 Barcelona, Spain 
         \and
              Institut f\"ur Astrophysik, Georg-August-Universit\"at,
              Friedrich-Hund-Platz 1, D-37077 G\"ottingen, Germany
        \and
              Centro de Astrobiolog\'ia (CSIC-INTA), ESAC,
              Camino Bajo del Castillo s/n, E-28692 Villanueva de la Ca\~nada, Madrid, Spain
        \and
              Landessternwarte, Zentrum f\"ur Astronomie der Universt\"at Heidelberg,
              K\"onigstuhl 12, D-69117 Heidelberg, Germany
        \and
              Instituto de Astrof\'isica de Andaluc\'ia (IAA-CSIC),
              Glorieta de la Astronom\'ia s/n, E-18008 Granada, Spain
        \and
              Hamburger Sternwarte,
              Gojenbergsweg 112, D-21029 Hamburg, Germany
        \and
              Homer L. Dodge Department of Physics and Astronomy, University of Oklahoma, 440 West Brooks Street, Norman, OK 73019, USA
        \and
              Instituto de Astrof\'isica de Canarias,
              V\'ia L\'actea s/n, E-38205 La Laguna, Tenerife, Spain
         \and
              Departamento de Astrof\'isica, Universidad de La Laguna, E-38026 La Laguna, Tenerife, Spain
        \and
              Spanish Virtual Observatory 
        \and
              Centro Astron\'onomico Hispano Alem\'an, Observatorio de Calar Alto, Sierra de los Filabres, E-04550 G\'ergal, Spain
        \and
              Th\"uringer Landesstenwarte Tautenburg,
              Sternwarte 5, D-07778 Tautenburg, Germany
        \and
              Max-Planck-Institut f\"ur Astronomie,
              K\"onigstuhl 17, D-69117 Heidelberg, Germany
        \and
              Departamento de F\'{i}sica de la Tierra y Astrof\'{i}sica and IPARCOS-UCM (Instituto de F\'{i}sica de Part\'{i}culas y del Cosmos de la UCM), Facultad de Ciencias F\'{i}sicas, Universidad Complutense de Madrid, E-28040 Madrid, Spain
             }

   \date{Received 20 Apr 2020 / Accepted 20 Jun 2020} 

 
  \abstract
   {Variability caused by stellar activity represents a challenge to the discovery and characterization of terrestrial exoplanets and complicates the interpretation of atmospheric planetary signals.}
   {We aim to use a detailed modeling tool to reproduce the effect of active regions on radial velocity measurements, which aids the identification of the key parameters that have an impact on the induced variability.}
   {We analyzed the effect of stellar activity on radial velocities as a function of wavelength by simulating the impact of the properties of spots, shifts induced by convective motions, and rotation. We focused our modeling effort on the active star YZ\,CMi (GJ\,285), which was photometrically and spectroscopically monitored with CARMENES and the Telescopi Joan Or\'o.}
   {We demonstrate that radial velocity curves at different wavelengths yield determinations of key properties of active regions, including spot-filling factor, temperature contrast, and location, thus solving the degeneracy between them. Most notably, our model is also sensitive to convective motions. Results indicate a reduced convective shift for M dwarfs when compared to solar-type stars (in agreement with theoretical extrapolations) and points to a small global convective redshift instead of blueshift.}
   {Using a novel approach based on simultaneous chromatic radial velocities and light curves, we can set strong constraints on stellar activity, including an elusive parameter such as the net convective motion effect.} 
   \keywords{convection -- stars: activity -- stars: low-mass -- stars:starspots -- techniques: radial velocities} 

   \maketitle

%

\section{Introduction} \label{sec:introduction}

Late-type dwarf stars are a strong focus of attention in the search for Earth-like planets using radial velocities (RVs), because the amplitude of their signals is larger than for solar-type stars \citep[e.g.,][]{Marcy1998,Bonfils2013,Perger2017}. However, with state-of-the-art instruments reaching  $\sim$1\,m\,s$^{-1}$ or even better uncertainties, it is not instrumental precision but astrophysical jitter that effectively limits the detection of exoplanets. This is particularly the case of some M-dwarf stars, because their intrinsic high level of magnetic activity causes the appearance of photospheric features such as stellar spots producing signals associated with the rotation period, which can hamper exoplanet detection efforts \citep{Benedict1993}. Several works discussing controversial exoplanet detections due  to the stellar intrinsic jitter or spurious signals in RV curves caused by data treatment have been published \citep[see, e.g.,][]{Robertson2014,Robertson2015,Rajpaul2015}. This reflects the difficulty of disentangling exoplanet signals from stellar jitter, even with the aid of stellar activity indices derived from spectroscopic data. However, it is also obvious that precise modeling of spot properties can help to disentangle and correct for stellar activity effects, thus enabling the detection of exoplanet signals that would otherwise be hidden within the stellar RV jitter.

A number of studies have shown that RV variability caused by stellar spots is wavelength dependent, because the flux contrast of cold spots and hot faculae is smaller toward the infrared \citep[][]{Desort2007,Reiners2010}. Thus, in principle, measurements obtained at different spectral bands can be used to correct for intrinsic stellar RV jitter. However, the picture may be much more complicated because the effects of limb darkening, convection, and magnetic field, for instance, can cause both amplitude and phase differences between radial velocities derived from spectra at different wavelengths. On the other hand, these differential wavelength-dependent effects can be used to constrain the properties of stellar active regions.

The influence of stellar heterogeneities on RVs has been thoroughly studied in the past years by modeling the stars with surface elements at different effective temperatures representing the immaculate photosphere, spots, and faculae \citep[see, e.g.,][]{Saar1997,Hatzes2002,Lanza2007,Boisse2012}. Photometric and RV variability due to stellar spots are then computed by disk-integrating the spectra corresponding to each surface element at different rotation phases. This approach only takes into account the effect of the flux dependence with effective temperature. 

It is well known that the presence of magnetic fields also changes the properties of the convective layer of dwarf stars \citep{Title1987,Hanslmeier1991}. Convective cells cause significant effects on RVs, as motions produce net shifts and distortions on spectral lines \citep{Dravins1999,Livingston1999}. For example, \cite{Gray2009} and \cite{Meunier2017} showed that RVs derived from different spectral lines depend on the depth of the line, which also affects the shape and absolute shift of the cross-correlation function bisectors. This makes the estimation of stellar convective motions strongly dependent on the spectral lines employed and on the temperature distribution of surface elements at the time of observation. This is true even for the Sun, for which measures of convective (blue)shift range from 200\,m\,s$^{-1}$ to 500\,m\,s$^{-1}$ \citep{Meunier2010,Lanza2010}. \cite{Meunier2017} used the differential Doppler displacement between the spectral lines and their dependence on the line depth to compute the convective shift for a sample of G0 to K2 main-sequence stars, and found that the absolute value of the convective shift decreases toward cooler stars. Hydrodynamic numerical simulations performed by \cite{Allende2013} showed a similar dependence of the convective blueshift on spectral type. However, there appear to be no direct measurements of the convective shift for M dwarfs, and it is not yet certain if the decreasing trend of the convective shift continues toward late-type stars, or even if it becomes convective redshift, as some indirect measurements suggest \citep{Kurster2003}.

Regarding general active region properties, measurements of spot sizes and locations have been made using the Doppler imaging technique \citep{Vogt1983}, obtaining filling factors that reach $\sim$10\,\% of the stellar surface \citep{Strassmeier2009}, or even larger, as were found studying magnetic regions with Zeeman-Doppler imaging \citep{Donati1997,Morin2008}. Other techniques such as light-curve inversion \citep{Messina1999,Berdyugina2002} can be used to retrieve spot sizes and contrast temperatures, although with this method the determination of spot temperature is strongly correlated with spot size. However, this limitation can be overcome by analyzing light curves covering a wide range of photometric bands \citep{Mallon2018,Rosich2020}. Modeling of photometric variations of late-type active stars has revealed that cool starspots are often quite large, covering up to 20\,\% of the stellar surface \citep{Berdyugina2005}. Regarding spot contrast temperatures, \cite{Berdyugina2005} gave a representative sample of measurements suggesting values decreasing from $\sim$2000\,K for late F- and G-type stars to $\sim$200\,K for mid-M dwarfs. It should be noted that this work used a very limited and heterogeneous sample of stars, particularly M dwarfs, and combined spot temperatures determined from different methods, some of them prone to systematic biases and degeneracies.

In this paper, we demonstrate that the \texttt{StarSim} stellar activity model code \citep{Herrero2016,Rosich2020} 
can provide stringent constraints on the properties of spots. In particular, a simultaneous fit to light and RV curves for several wavelength bands allows us to break the degeneracy between the spot coverage area and the temperature contrast, as well as to analyze the convective shift, providing a novel approach to measure this parameter. The chromatic index (CRX) introduced by \cite{Zechmeister2018} in the context of the CARMENES survey\footnote{\tt \url{http://carmenes.caha.es}} \citep[Calar Alto high-Resolution search for M dwarfs with Exo-earths with Near-infrared and optical \'Echelle Spectrographs;][]{Quirrenbach_2016,Quirrenbach_2018}, which measures the dependence of the radial velocity on wavelength, is ideally suited to the study of stellar activity effects. This index is defined as the slope of a linear fit to the RV as a function of the central wavelength logarithm of each order in a cross-dispersed \'echelle spectrum at each time step. Its use as an activity indicator relies on the wavelength dependence of photospheric heterogeneities because of the temperature contrast \citep{Barnes2011, Jeffers2014}. This dependence generally causes a decrease in the activity-induced RV variations toward the reddest orders \citep{Desort2007}, although strong magnetic fields may also have a large impact at longer wavelengths due to the Zeeman effect \citep{Reiners2013,Shulyak2019}. 

In Sect.\,\ref{sec:modelling}, we describe the model that we used to simulate time series of a rotating star with active regions, and we analyze which photospheric parameters play an important role. In Sect.\,\ref{sec:estimation1}, we present the spectroscopic and photometric observations of the active star YZ\,CMi and the results of the modeling of the spot parameters. Finally, in Sect.\,\ref{sec:discussion}, we compare our results with previous parameter estimations using other methods and present our conclusions.

\section{Modeling a spotted rotating star with \texttt{StarSim}} \label{sec:modelling}

\subsection{{\rm \texttt{The StarSim}} model}\label{sub:starsim}

\texttt{StarSim} is a sophisticated model used for simulating the effects of stellar spots on light and radial velocity curves. It allows us to generate precise synthetic photometric and spectroscopic time-series data of a spotted rotating photosphere. We briefly introduce the key aspects of the model here, but we refer the reader to \cite{Herrero2016} for a detailed description. The model is based on the integration of the spectral contribution of a fine grid of surface elements. 

Synthetic PHOENIX spectra of different temperatures \citep{Husser2013} are assigned to each of the surface elements, with a different temperature depending on the properties of the region (quiet photosphere, spot, or facula), and Doppler-shifted according to the projected velocity of the surface element. Photometric light curves or RVs are then computed by integrating all surface elements. To speed up the computation of RVs and other spectral indices related to the cross-correlation function (hereafter, CCF), \texttt{StarSim} initially generates the CCFs produced from the spectrum of a single photosphere, spot, and facular element, and then integrates the entire visible surface using CCFs instead of spectra. A slow rotator template or a user-defined mask of spectral lines can be used to calculate the CCFs. Then, the contribution from each surface element is adjusted for the limb darkening computed from Kurucz ATLAS9 models \citep{Kurucz2017} at the specific angle with respect to the line of sight. \texttt{StarSim} defines the facular elements as circular regions around spots, whose area is controlled by the facula-to-spot area ratio. To properly account for the center-to-limb variations of convection effects in active regions, \texttt{StarSim} subtracts the bisector of the CCFs computed from the original PHOENIX spectra and then adds the bisector computed from CIFIST 3D models of a Sun-like star \citep[][]{Ludwig2009}. The program also allows the adding of an arbitrary extra convective shift (blueshift or redshift) to the bisector of the CCFs corresponding to regions covered with spots or faculae. Finally, RV values are determined from the CCF at each epoch, which takes into account the distribution of spots and faculae and their properties, by fitting a Gaussian function.

Several stellar input parameters can be set in \texttt{StarSim}, such as the effective temperature of the star, the spot temperature, the position, size, and number of active regions, the convective shift, the stellar rotation period, the radius of the star, its surface gravity, and the inclination of the stellar spin axis with respect to the line of sight. Furthermore, one can select the wavelength range and compute time-series data of the photometry, RV, and CCF parameters. 

To calculate the CRX parameter, we simulated RV curves for different wavelength ranges and measured their wavelength dependence. For this study, we used wavelength ranges matching those of the CARMENES visual channel \'echelle orders \citep[i.e., 61 orders from 520 to 960\,nm,][]{Quirrenbach_2016,Quirrenbach_2018} to be consistent with our observational data. However, following \cite{Zechmeister2018}, to compute the simulated CRX, we used the 40 orders (\'echelle orders 68 to 108) where the signal-to-noise ratio of the spectra is the highest, and thus observational RVs have lower uncertainties. The CRX is then computed as the slope of a linear fit to the RV as a function of the logarithm of the central wavelength of each order.

\subsection{Dependence on photospheric parameters}\label{sub:parameters}

To study the dependence of the RV and CRX time series on the properties of active regions, we ran several simulations considering a rotating spotted star. For simplicity, we assumed a single circular spot on the photosphere, and tested different values for the ratio of stellar surface covered by the spot (filling factor, \textit{ff}), the temperature difference between the photosphere and the spot ($\Delta T \equiv T_{\rm ph} - T_{\rm sp}$), and the convective shift ($CS$). As explained in Sect.\,\ref{sub:starsim}, \texttt{StarSim} introduces the influence of convective motions by adding a solar-like bisector to the CCF of each surface element. Thus, the solar convective blueshift is used as a reference. As an estimation of its absolute value, in our case a solar blueshift of 300 m\,s$^{-1}$ \citep{Dravins1981,Cavallini1985,Lohner2018} was subtracted from all $CS$ values shown in this work. Therefore, the absolute values of $CS$ given in this work have a dependence on the precise determination of the convective blueshift from CIFIST 3D models. 

Stellar parameters such as effective temperature, surface gravity, and metallicity also have an impact on the RV and CRX time series due to their effects on stellar spectra. Besides this, the rotation of the star, its inclination, the latitude of the spot, and the presence of faculae do also play a role. However, these stellar parameters can be determined from fits to high-resolution spectra or through independent data. For this reason, we focus only on the impact of the properties of spots on RV and CRX time series. 

For our simulations, we considered the case of a mid-M dwarf star approximately matching the properties of YZ\,CMi, which we use as a case study to compare with real observations later in Sect.\,\ref{sec:estimation1}. We fixed the stellar photospheric effective temperature to a value of 3100\,K, the stellar surface gravity to $\log g=5.0$, and we adopted solar metallicity. We assume an inclination of the rotation axis of $i=90$\,deg (edge-on), and we introduce a spot on the equator, with no evolution and no differential rotation. The facula-to-spot area ratio was set to zero is as commonly found for active M-dwarf stars \citep{Herrero2016,Mallon2018,Rosich2020}, and also as suggested by convection in magnetohydrodynamic simulations \citep{Beeck2011,Beeck2015}.

\begin{figure*}[t]
\centering
\includegraphics[width=0.32\textwidth]{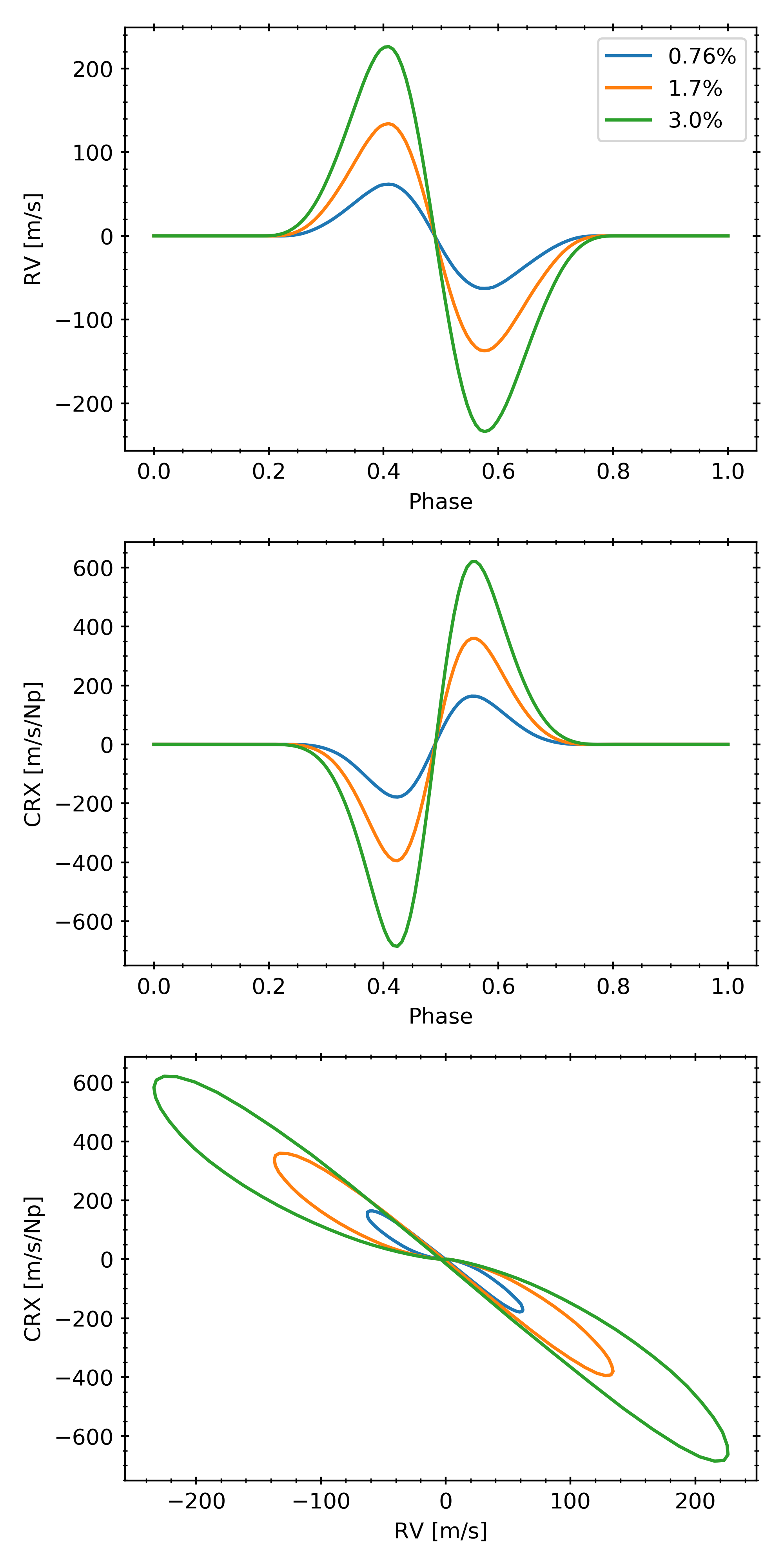}
\includegraphics[width=0.32\textwidth]{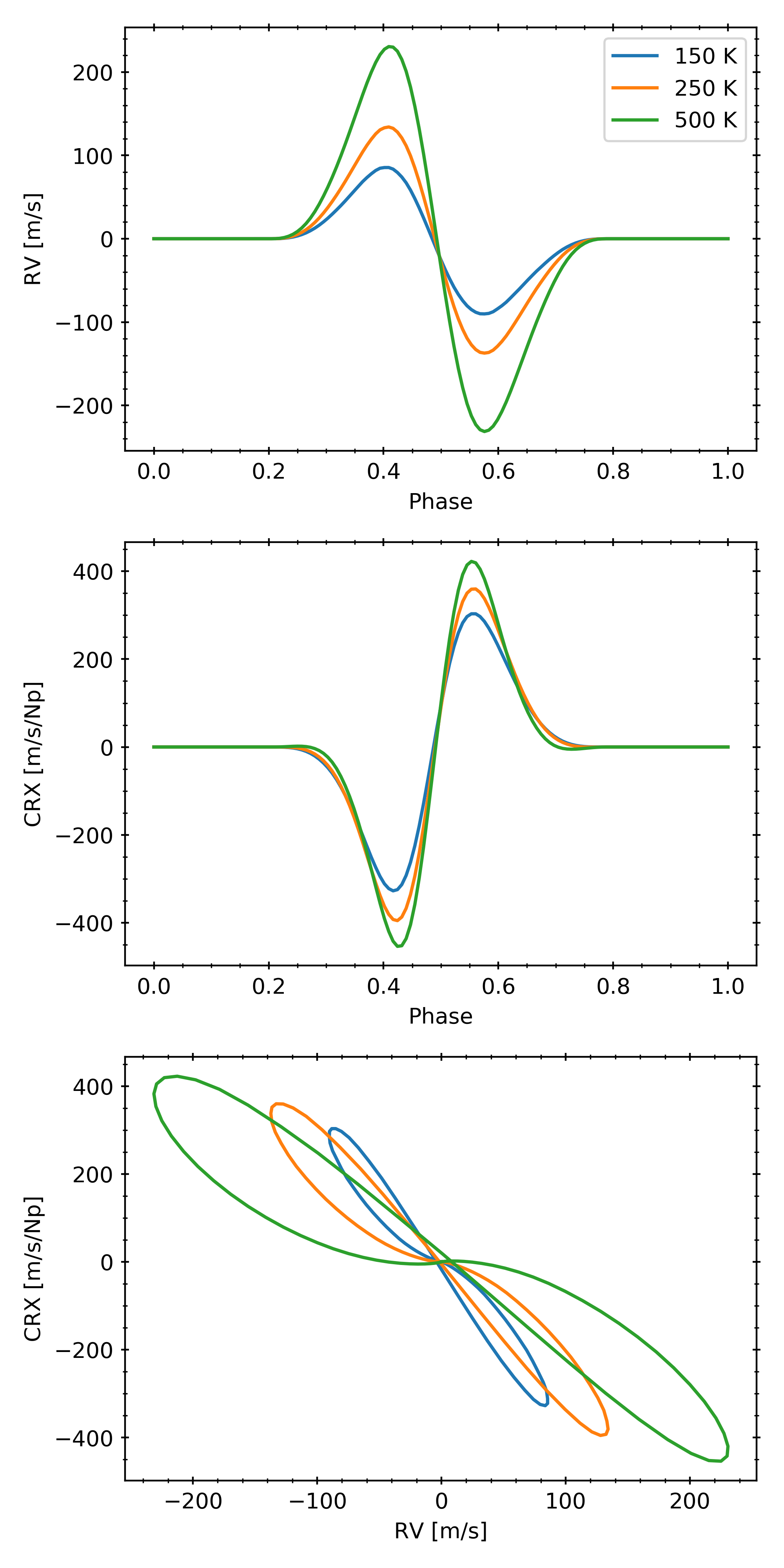}
\includegraphics[width=0.32\textwidth]{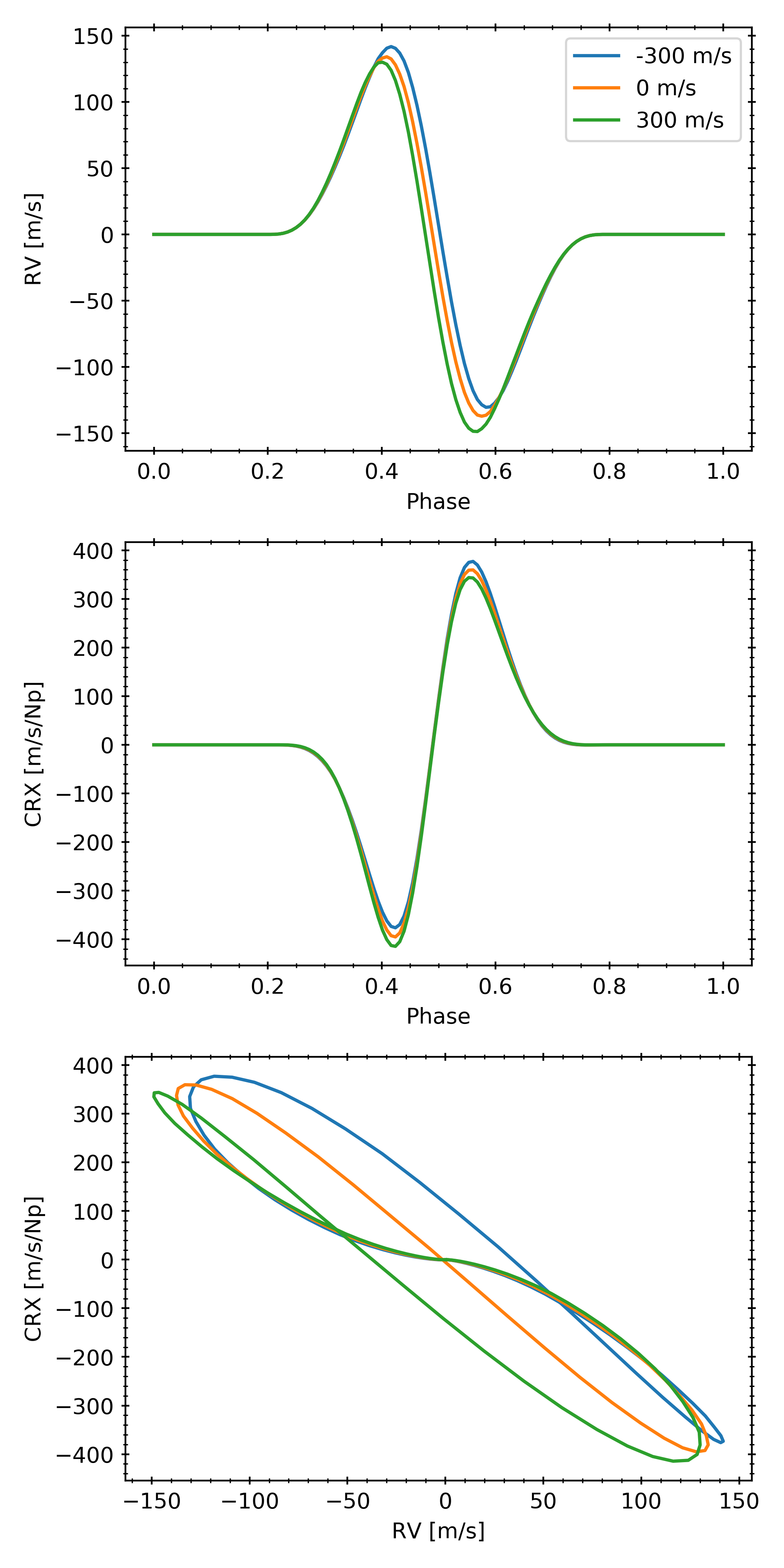}

\caption{Effect of properties of stellar spots on the RV (\textit{top panels}) and CRX (\textit{middle panels}) phase curves and on the RV-CRX correlation (\textit{bottom panels}) for a mid-M dwarf. Simulations of varying \textit{ff} ({\it left panels}), $\Delta T$ ({\it middle panels}), and $CS$ ({\it right panels}) are shown while keeping the other parameters constant to a set of reference values. These reference values are 1.7\,\% for the filling factor, 250\,K for $\Delta T$, and 0\,m\,s$^{-1}$ for $CS$. We note the different vertical scale in each plot, and that the horizontal axes are different in the bottom panels.}
          \label{example}%
\end{figure*}

Figure~\ref{example} illustrates the outcome of the simulations as a function of \textit{ff}, $\Delta T$ and $CS$ for RV and CRX. The bottom panels show that the CRX is strongly anti-correlated with the RV. As other authors have previously shown for the case of the bisector span \citep[see, e.g.,][]{Boisse2011,Figueira2013}, the CRX-RV anti-correlation is not a straight line, but it shows a lemniscate-like structure. The left panels in Fig.\,\ref{example} show that varying \textit{ff} mainly produces a change in the amplitude of both the RV and CRX time series, which in turn changes the scale of the correlation but preserving its slope and shape. The middle panels display the effect caused by different values of $\Delta T$. This parameter has a much higher impact on the amplitude of the RV curve than on the CRX, producing a significant change in the slope of the correlation. Finally, $CS$ changes the phase at which the RV peaks, and only slightly alters the asymmetry between maximum and minimum of both the RV and CRX. This change in phase does not produce any change in the scale and slope of the RV-CRX correlation, but it has a significant imprint on its shape.

The different effects of stellar spot parameters on RV and CRX data can be understood by considering the main sources of variability, which are: \emph{($i$)} the flux effect caused by the contrast between inhomogeneities and the immaculate photosphere \citep{Dumusque2014}, and \emph{($ii$)} the inhibition of convection in active regions \citep{Dravins1981,Stein1992,Chabrier2007}. Both effects cause periodic variations of the measured RV of spotted stars. However, their dependence on rotation phase (or time) is different \citep[see, e.g., Fig.\,7 in][]{Herrero2016}. The flux effect vanishes when the spot is facing the observer (central phase) because the spot covers equivalent surface areas moving toward and away from the observer, hence, the net effect is canceled. Besides this, the flux effect is anti-symmetric with respect to this point, meaning it causes an RV maximum when the spot crosses the hemisphere approaching the observer, and a minimum on the other side. On the other hand, the convection effect is maximal at the central phase, when the projected filling factor of the spot is largest, and the RV variability is symmetric with respect to this point. The difference between both effects produces a phase shift of the RV peaks for different wavelengths, causing the $\infty$ shape of the CRX-RV correlation, which becomes asymmetric depending on the parameters. The top-right panel in Fig.\,\ref{example} shows that the phase shifts of RV time series are dominated by $CS$, while CRX remains almost unaffected (middle-right panel). The peak-to-peak amplitudes of both observables are mainly determined by the size and the temperature of the spot. This makes the analysis of chromatic radial velocities a unique tool to constrain $CS$, by simultaneously fitting the peak-to-peak amplitude and the phase shift of RV and CRX time series.

\section{Fitting chromatic radial velocities of YZ\,CMi} \label{sec:estimation1}

\subsection{The active star YZ\,CMi} \label{sub:yzcmi}

The star YZ\,CMi (GJ\,285) is a young star belonging to the $\beta$ Pictoris moving group \citep[][and references therein]{Alonso2015}. Its main properties are listed in Table\,\ref{tab:props}. \cite{VanMaanen1945} first announced the BY\,Dra-type photometric variability of the star, which was also classified as a flaring star by \cite{Lippincott1952}, showing UV\,Cet-type flares. Several flaring events were reported later \citep[][]{Andrews1966,Sanwal1976,Zhilyaev2011}. A rotation period of 2.77\,d from photometric variations was reported by \cite{Chugainov1974} and confirmed by \cite{Pettersen1983}. Recent estimations of the rotation period yielded very similar values \citep[$2.7758\pm0.0006$\,d and $2.776\pm0.010$\,d,][respectively]{Morin2008,DiezAlonso2019}. \cite{Bondar2019} found an activity cycle of $27.5\pm2.0$\,yr, estimated from more than 80 years of archival photometric observations, with peak-to-peak variations of 0.2--0.3\,mag. Estimations of the effective temperature of YZ\,CMi range from 3045\,K \citep[from synthetic spectra fitting;][]{Rojas2012} to 3600\,K \citep[from spectral color indices;][]{Zboril2003}.

The variability of the star has been extensively studied. The large color excess of YZ\,CMi indicates the presence of cool spots covering a large fraction of its surface. \cite{Zboril2003} computed spot solutions from light curves taken in different seasons, and found a typical spot coverage between 10\,\% and 25\,\% of the surface, resulting from a single spot at a co-latitude $\sim$15--35\,deg from the pole and temperature $\sim$500\,K cooler than the surrounding photosphere, all assuming an inclination ranging from 60\,deg to 75\,deg. Similarly, \cite{Alekseev2017} used 11 epochs of observations over 30 years of broad-band photometry to compute the spottedness of YZ\,CMi. The authors assumed an effective temperature of 3300\,K and an inclination of $i=60$\,deg, obtaining two belts of spots $210\pm70$\,K cooler than the photosphere, which were mainly located at latitudes of 12--15\,deg, and covering up to 38\,\% of the stellar surface. Using Zeeman-Doppler imaging and 25 spectropolarimetric observations taken between 2007 and 2008, \cite{Morin2008} found that the visible pole is covered by a large spot, with strong axisymmetry in the magnetic energy modes, hinting toward a non-visible spot at the other hemisphere. \cite{Morin2008} also inferred negligible differential rotation ($0.0\pm1.8$\,mrad\,d$^{-1}$).

We homogeneously derived the bolometric luminosity $L$ and effective temperature $T_{\rm eff}$ of YZ\,CMi using broadband photometry in 17 passbands, from optical blue $B_T$ \citep[Tycho-2,][]{Hog2000} to mid-infrared $W4$ \citep[AllWISE,][]{Cutri2013}. None of the measurements used, especially at the bluest passbands, seemed to be affected by strong flaring activity. To determine $L$ and $T_{\rm eff}$, we performed fits to the spectral energy distribution (SED), employing the BT-Settl CIFIST theoretical grid of models \citep{Baraffe2015} and the Virtual Observatory SED Analyzer \citep[VOSA,][]{Bayo2008}. The procedure is described in detail in \citet{Cifuentes2020}. We took into account the long-term photometric variability of YZ\,CMi in the blue optical by re-running the SED fitting at the brightness maximum and minimum. For that, we were very conservative and used the largest reported variability amplitude of 0.3\,mag in the $B$ band \citep{Bondar2018}, over a scale of decades, to recompute extreme values of bolometric luminosity. We approximately extrapolated such amplitudes to 0.2\,mag in the red optical, 0.1\,mag in the near infrared, and 0.05\,mag in the mid infrared, as observed in multiband photometric monitoring of very active M dwarfs \citep[e.g.,][]{Caballero2006}. The corresponding $L$ and $T_{\rm eff}$ uncertainties of YZ\,CMi are thus larger than for invariable field M dwarfs of similar brightness. The radius was calculated using Stefan-Boltzmann's law following \cite{Schweitzer2019} and propagating the uncertainties in $L$ and $T_{\rm eff}$.

We estimated the inclination of the stellar spin axis from the radius, the rotation period and the projected velocity, which yielded $i=36^{+17}_{-14}$\,deg. The uncertainty was computed from $10^6$ random resamplings of the input parameters according to their quoted uncertainties. The lower inclination value that we find, contrary to other studies \citep{Zboril2003,Morin2008,Alekseev2017}, arises from the different projected velocity and stellar radius used. In particular, projected rotational velocities of 5.0 and 6.5\,km\,s$^{-1}$ \citep[][]{Delfosse1998,Reiners2007} and stellar radii of 0.30 and 0.37\,$R_{\odot}$ \citep{Delfosse2000,Pettersen1980} were reported. Based on CARMENES data, \cite{Reiners2018} reported $v \sin{i} = 4.0 \pm 1.5$\,km\,s$^{-1}$, and we find a stellar radius of 0.369\,R$_{\odot}$ following \cite{Schweitzer2019}. The resulting parameters are listed in Table~\ref{tab:props}.

\begin{table}[t]
\centering
\caption{Basic properties of YZ\,CMi.}
\label{tab:props}
\begin{tabular}{lcl} 
\hline\hline
\noalign{\smallskip}
Parameters & Values  & Ref.\\
\noalign{\smallskip}
\hline
\noalign{\smallskip}
GJ  & 285   & GJ79 \\ 
Karmn & J07446+035  & AF15  \\ 
$\alpha$ (J2000) & 07:44:40.17&  {\it Gaia} DR2\\
$\delta$ (J2000) & +03:33:08.9 & {\it Gaia} DR2\\ 
$d$ [pc] & $5.9874\pm0.0021$ & {\it Gaia} DR2\\
$G$ [mag] & $9.6807\pm0.0010$ & {\it Gaia} DR2\\
$J$ [mag] & $6.581\pm0.024$ & 2MASS\\
Sp. type & M4.5\,V & PMSU\\ 
$T_{\rm eff}$ [K] & $3100\pm50$  & This work \\
$\log{g}$ [cgs]& $5.0\pm0.5$ & Lep13\\
$L_{\star}$ [10$^{-4}$\,$L_{\odot}$]& 113$^{+17}_{-14}$ & This work \\ 
$R_{\star}$ [$R_{\odot}$]& 0.369$^{+0.027}_{-0.055}$ & This work \\ 
pEW(H$\alpha$) [${\AA}$] &$-7.097\pm0.023$& Jef18\\ 
$v\sin{i}$ [km\,s$^{-1}$] & $4.0\pm1.5$ & Rei18\\
$P_{\rm rot}$ [d] & $2.776\pm0.010$ & DA19 \\
$i$ [deg] & $36^{+17}_{-14}$ & This work \\
\noalign{\smallskip}
\hline
\end{tabular}
\tablebib{ 
2MASS: \cite{Skrutskie2006}; 
AF15: \cite{Alonso2015}; 
DA19: \cite{DiezAlonso2019}; 
{\it Gaia} DR2: \cite{Gaia2016,gaia_2018}; 
GJ79: \cite{Gliese79};
Jef18: \cite{Jeffers2018}; 
Lep13: \cite{Lepine2013}; 
PMSU: \cite{hawley_1996}; 
Rei18: \cite{Reiners2018}. 
}
\end{table}

The selection of this target for analysis of its stellar activity properties was based on two main points: \emph{(i)} it is one of the M-dwarf stars with the strongest activity-induced RV variability within the CARMENES survey, with peak-to-peak amplitudes of $\sim$300\,m\,s$^{-1}$, and \emph{(ii)} it is also one of the stars with the strongest RV-CRX correlation  \citep{Talor2018}, which could be indicative of large spots.

\subsection{CARMENES observations} \label{sub:obs}

\begin{figure}[t]
\centering
\includegraphics[width=\columnwidth]{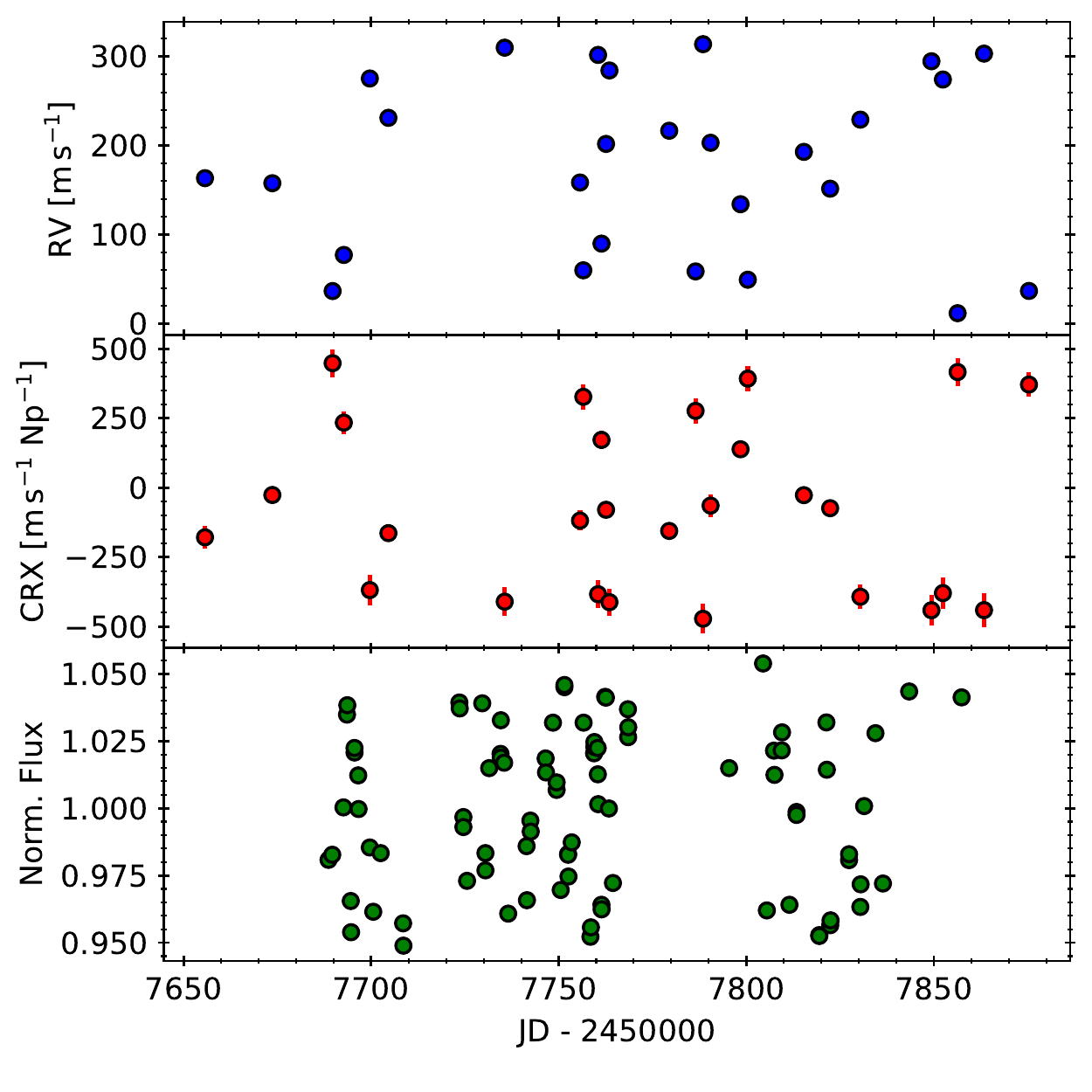}
\caption{RV (\textit{top}), CRX (\textit{middle}), and photometric (\textit{bottom}) time series of YZ\,CMi obtained with CARMENES and TJO.} 
          \label{timeseries}%
\end{figure}

YZ\,CMi is one of the targets of the CARMENES survey \citep{Reiners2018}. CARMENES is a double-channel spectrometer of which the aim is the discovery of exoplanets orbiting M-dwarf stars. It obtains high-resolution spectra of stars simultaneously in two different wavelength ranges, one in the red-visible and another one in the near-infrared, covering from 520 to 960\,nm and from 960 to 1710\,nm, respectively. Besides radial velocities and some commonly used spectral indices, the CARMENES {\tt serval} pipeline \cite[][]{Zechmeister2018} also computes the CRX. YZ\,CMi was observed between 2016 and 2018 at 49 epochs. To mitigate the possible effects of spot evolution, an observational campaign to sample all rotational phases was conducted between September 2016 and May 2017, yielding a total of 27 valid observations, which are used in this work. Figure\,\ref{timeseries} displays the RV and CRX time series, and all used data are available in Table\,\ref{tab:data} in the Appendix. Figure\,\ref{phasefit} shows the same measurements, but phase folded to the stellar rotation period of 2.776\,d \citep{DiezAlonso2019}. RV and CRX are anti-correlated, as was expected given the results of the simulations in Sect.\,\ref{sub:parameters}.

\begin{figure}[t]
\centering
\includegraphics[width=\columnwidth]{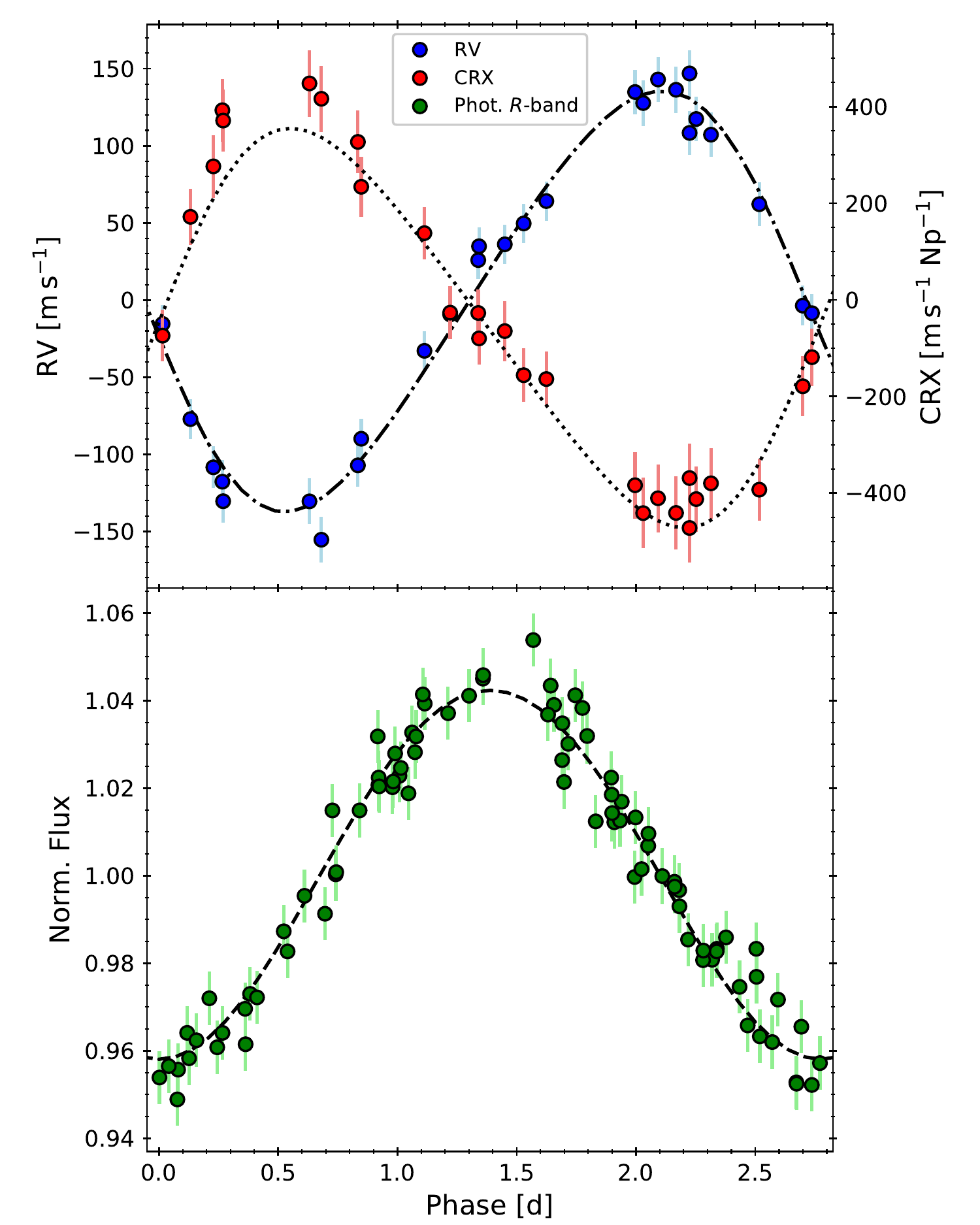}

\caption{Phase-folded RV (blue circles), CRX (red circles), and photometric (green circles) data obtained for YZ\,CMi. In the top panel, the left and right vertical axes correspond to RV and CRX, respectively. The best models that simultaneously fit the three datasets are shown as dash-dotted (RVs) and dotted (CRX) lines, and they correspond to the values found in Table \ref{tab:results}. The error bars correspond to the quadrature addition between observational uncertainties and the fit jitter parameter for each dataset.}
          \label{phasefit}%
\end{figure}

\subsection{Photometric data}

We also secured photometric monitoring of YZ\,CMi contemporaneously to RV observations. We obtained 460 photometric measurements using a Johnson~\textit{R} filter with the Telescopi Joan Or\'o (TJO), located at the Montsec Astronomical Observatory in Lleida, Spain. The TJO is a fully-robotic 0.8\,m Ritchey-Chr\'etien telescope. The photometric data were obtained with the MEIA2 instrument, an Andor 2\,k$\times$2\,k CCD camera with a plate scale of 0.36\,arcsec per pixel. We gathered these observations between October 2016 and April 2017. After rebinning the data with a cadence of 30\,min, we performed a 2.5\,$\sigma$-clipping to the residuals after a sinusoidal fit, in order to remove outliers due to flaring events, which resulted in a total of 89 photometric epochs. The bottom panels in Fig.\,\ref{timeseries} and Fig.\,\ref{phasefit} display the resulting photometric data. 

\subsection{Surface distribution of the active regions} \label{spot}

Before analyzing the impact of the size and the temperature of spots and the convective shift on the RV and CRX time series of YZ\,CMi, we needed to estimate the distribution of active regions on the stellar photosphere, particularly their latitude. Although light curves are not strongly sensitive to the latitude of active regions, including RV data in the analysis does solve this problem. This is especially true in the case of YZ\,CMi, for which a constraint on the inclination of the spin axis is available. For this purpose, we made use of the implementation of the inverse problem in \texttt{StarSim}, which allows the fitting of the spot distribution that best matches light or radial velocity curves \citep{Rosich2020}. This new implementation solves the inverse problem by using several small active elements on the star, which may be concentrated to reproduce spot groups. We used the stellar input parameters from Table \ref{tab:props}, and we estimated the latitude of active regions from the observed RV and photometric time series assuming different temperature contrast values in the range $\Delta T =$ 50--400\,K. We found that, while the size of active elements showed some correlation with their temperature contrast, their latitudes were always in the range 75--81\,deg. The inversion of the RVs using a model with 20 active surface elements resulted in all of them concentrated around the same region. For that reason, and to speed up and simplify the fitting procedure, we decided to further model the data with a larger single non-evolving circular spot located at a latitude of $\lambda_{\rm spot}=78$\,deg.

\subsection{Fit spot parameters} \label{sec:grid}

Assuming the spot model described above, we analyzed the impact of $\Delta T$, \textit{ff}, and $CS$ on the RV, CRX, and photometric time series of YZ\,CMi, and we determined the values that best fit all datasets simultaneously. The simulation of RV data using \texttt{StarSim} involves the computation of the CCF for each surface element of the star, a process that is computationally expensive. Therefore, we decided to fix the stellar parameters to those listed in Table\,\ref{tab:props}, and compared our simulations with the observed RV, CRX, and light curves on a grid of the $\Delta T$-\textit{ff}-$CS$ parameter space. To explore the parameter domain, the likelihood of the \texttt{StarSim} models given the observed datasets was first computed on a coarse grid, with 20\,K, $\sim$3.5\,\%, and 50\,m\,s$^{-1}$ steps in $\Delta T$, \textit{ff}, and $CS$, respectively. We subsequently defined a finer grid with step sizes divided by four to explore the regions with higher likelihood values. The likelihood at each grid point was computed by also considering an offset to the RVs ($\gamma_{\rm RV}$), and a global reference time shift to the RVs, CRX, and photometry ($\Delta t_{\rm ref}$), relating the central longitude of the star to the rotational phase. Finally, we also added jitter terms ($\sigma_{\rm RV}$, $\sigma_{\rm CRX}$, $\sigma_{\rm phot}$) to the RVs, CRX, and photometry, respectively, to account for the limitations of a single-spot model. These jitter terms are added in quadrature to the corresponding uncertainties to evaluate the likelihood function, as defined in \cite{Baluev2009}. The final model likelihood value reported is the sum of the individual likelihood values obtained for the RV, CRX, and photometry datasets.

\begin{figure*}[t]
  \centering
  \includegraphics[width=0.33\textwidth]{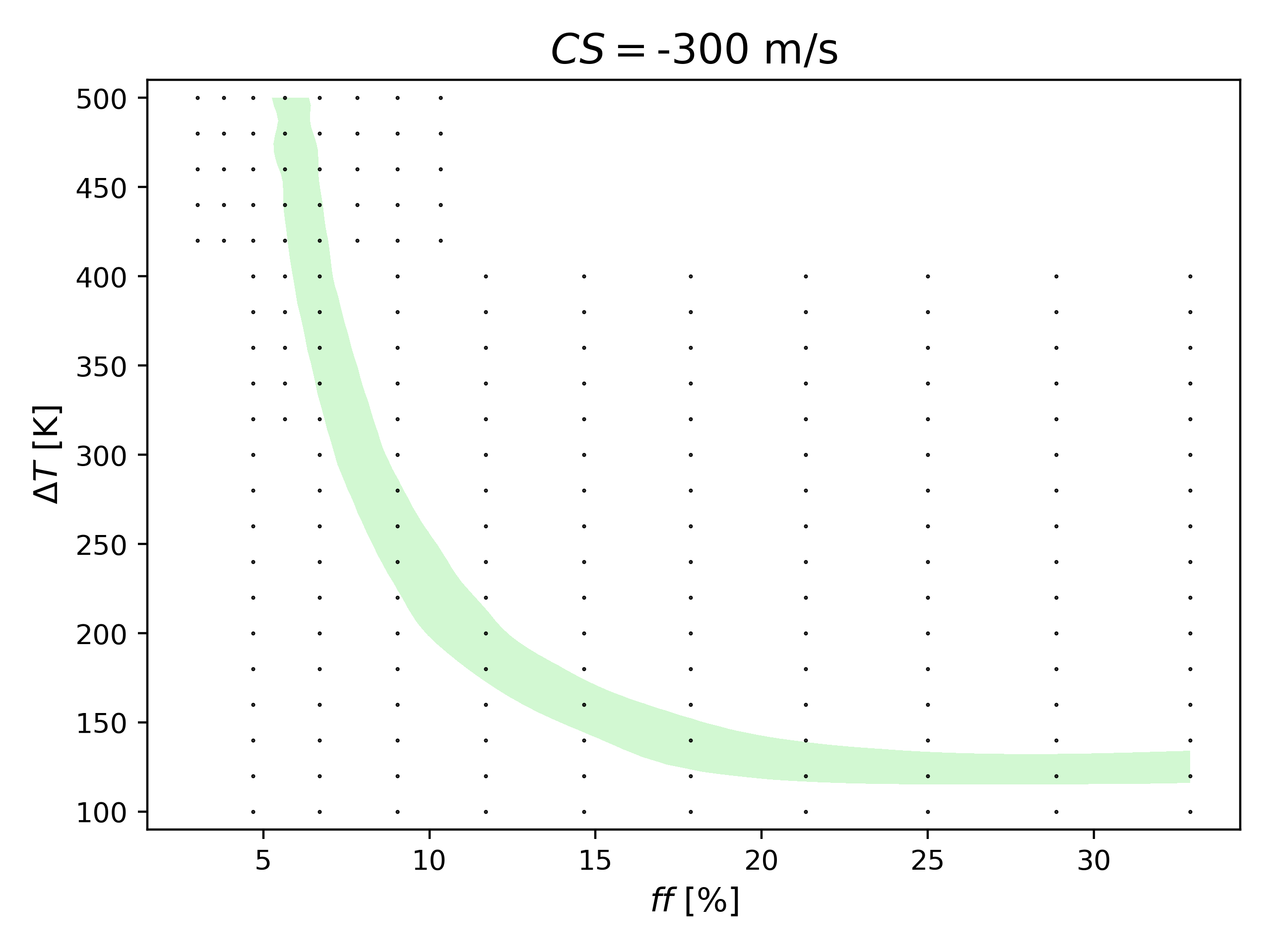}
  \includegraphics[width=0.33\textwidth]{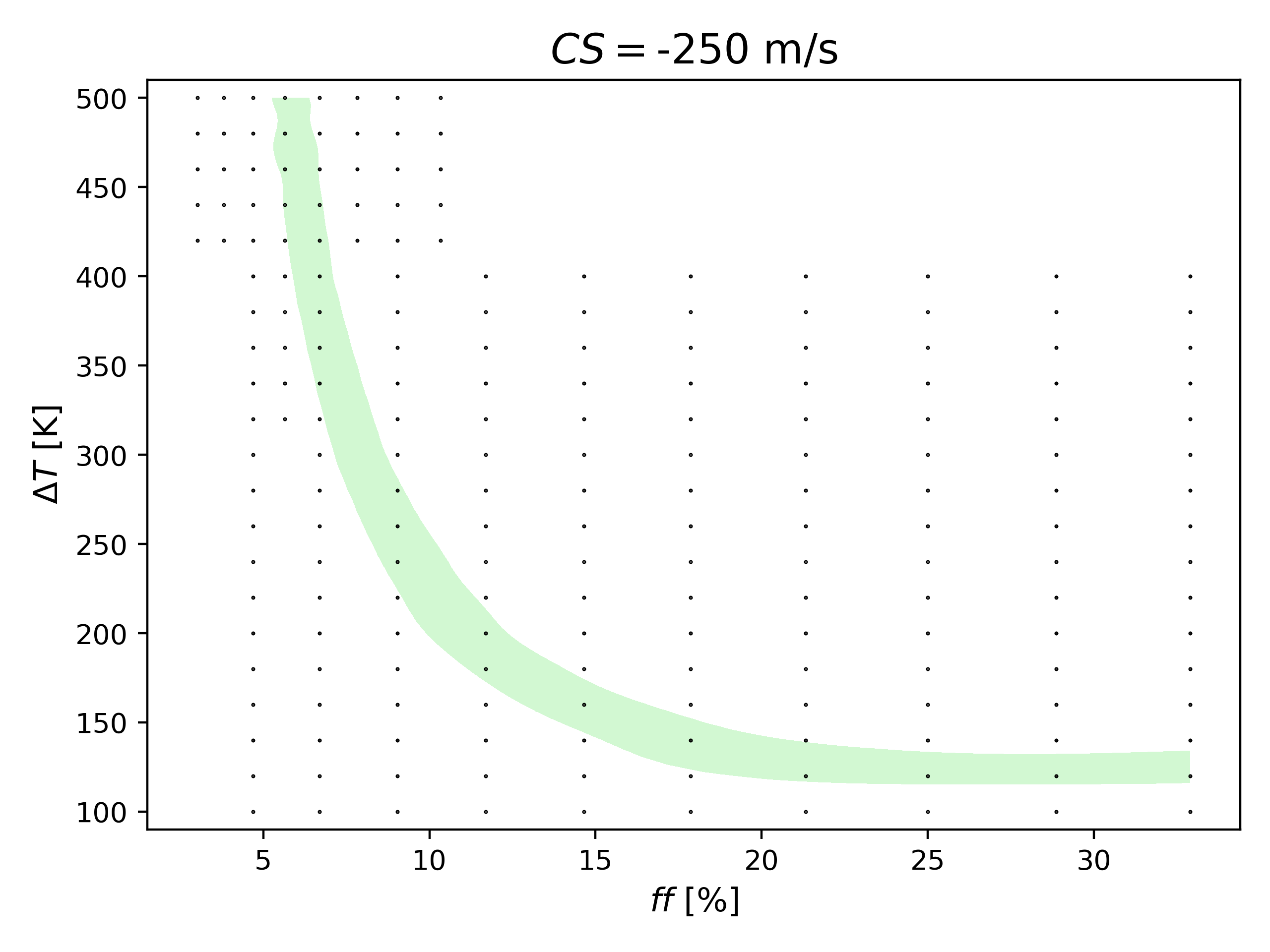}
  \includegraphics[width=0.33\textwidth]{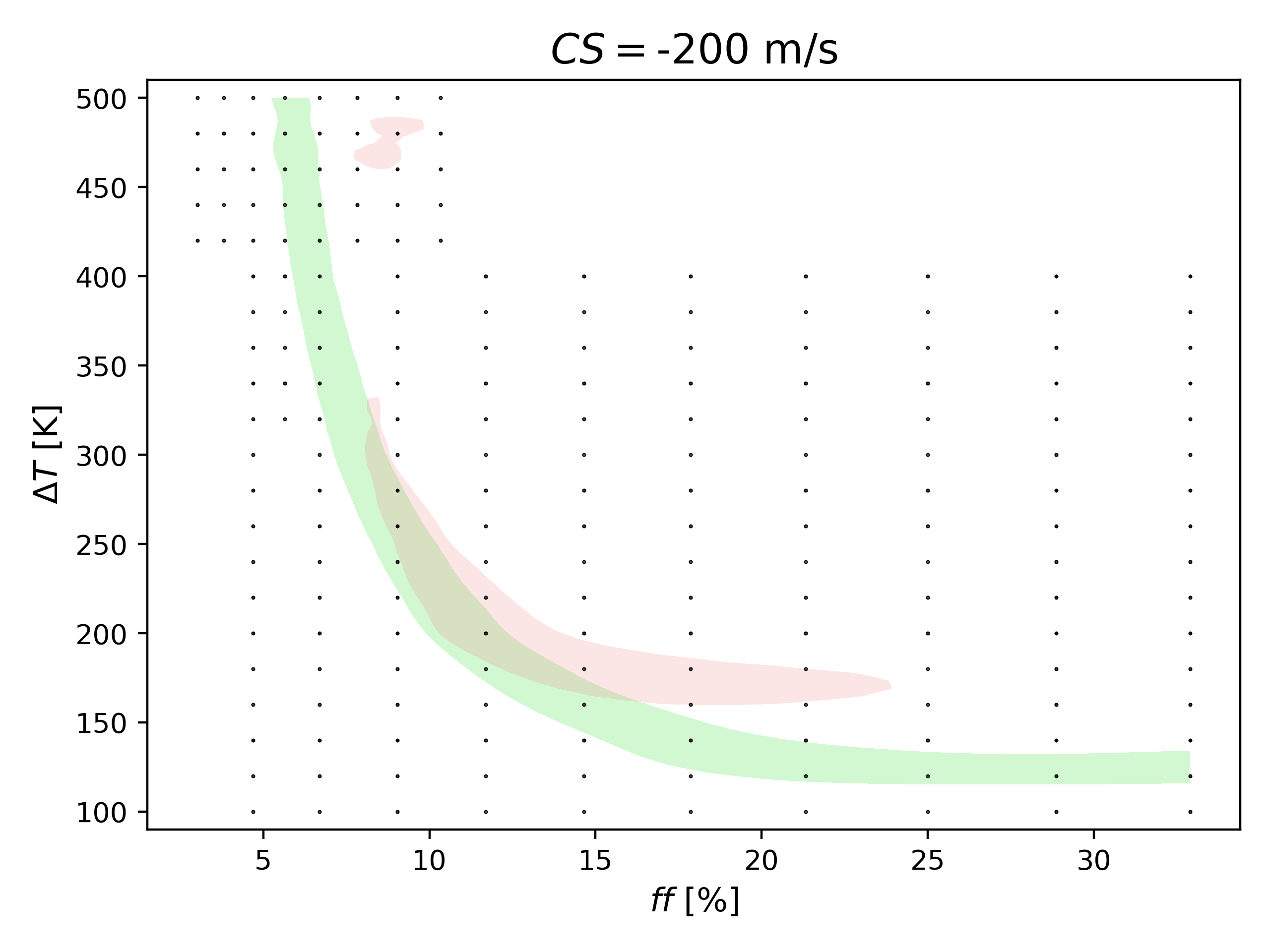}
  \includegraphics[width=0.33\textwidth]{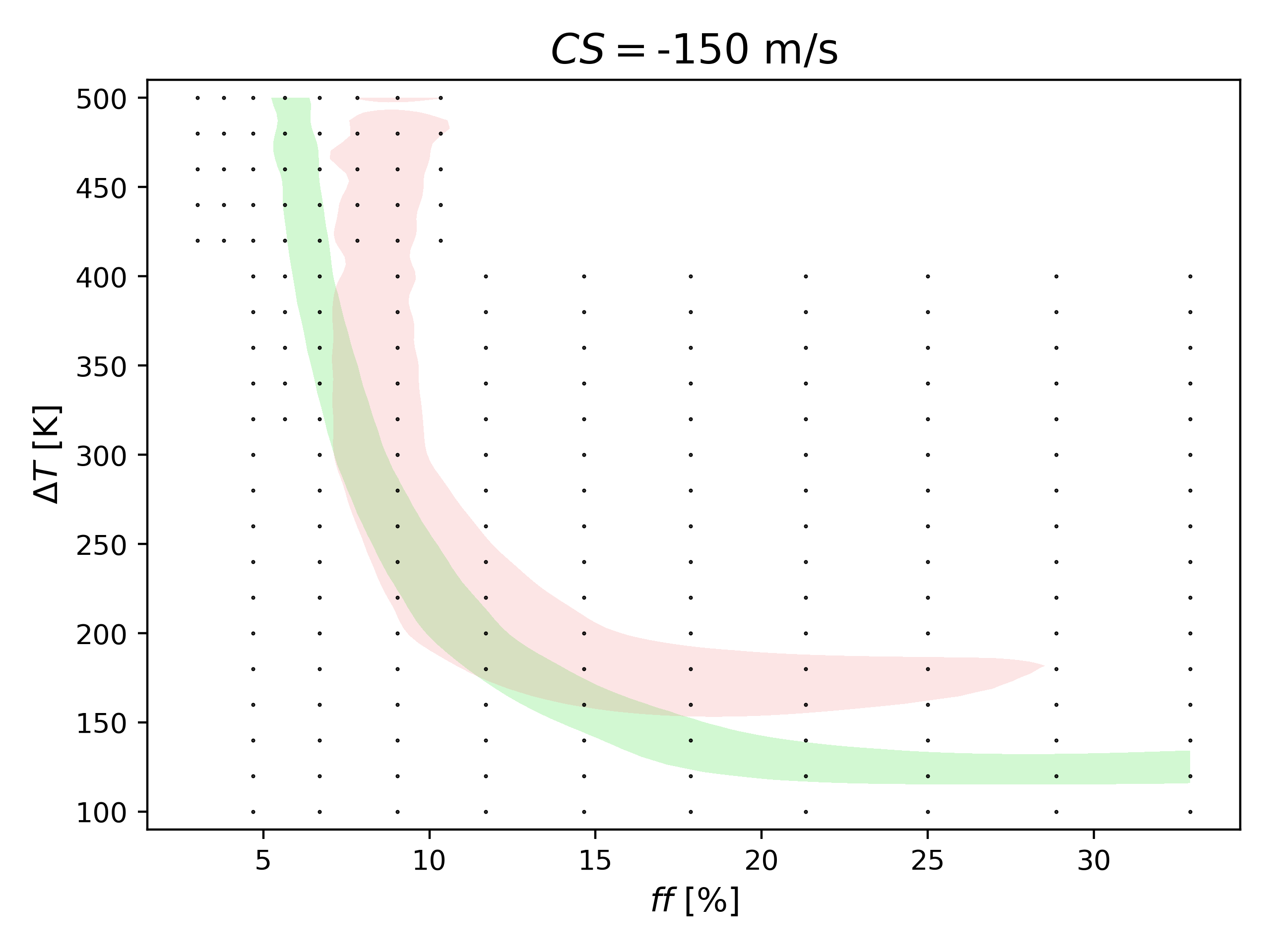}
  \includegraphics[width=0.33\textwidth]{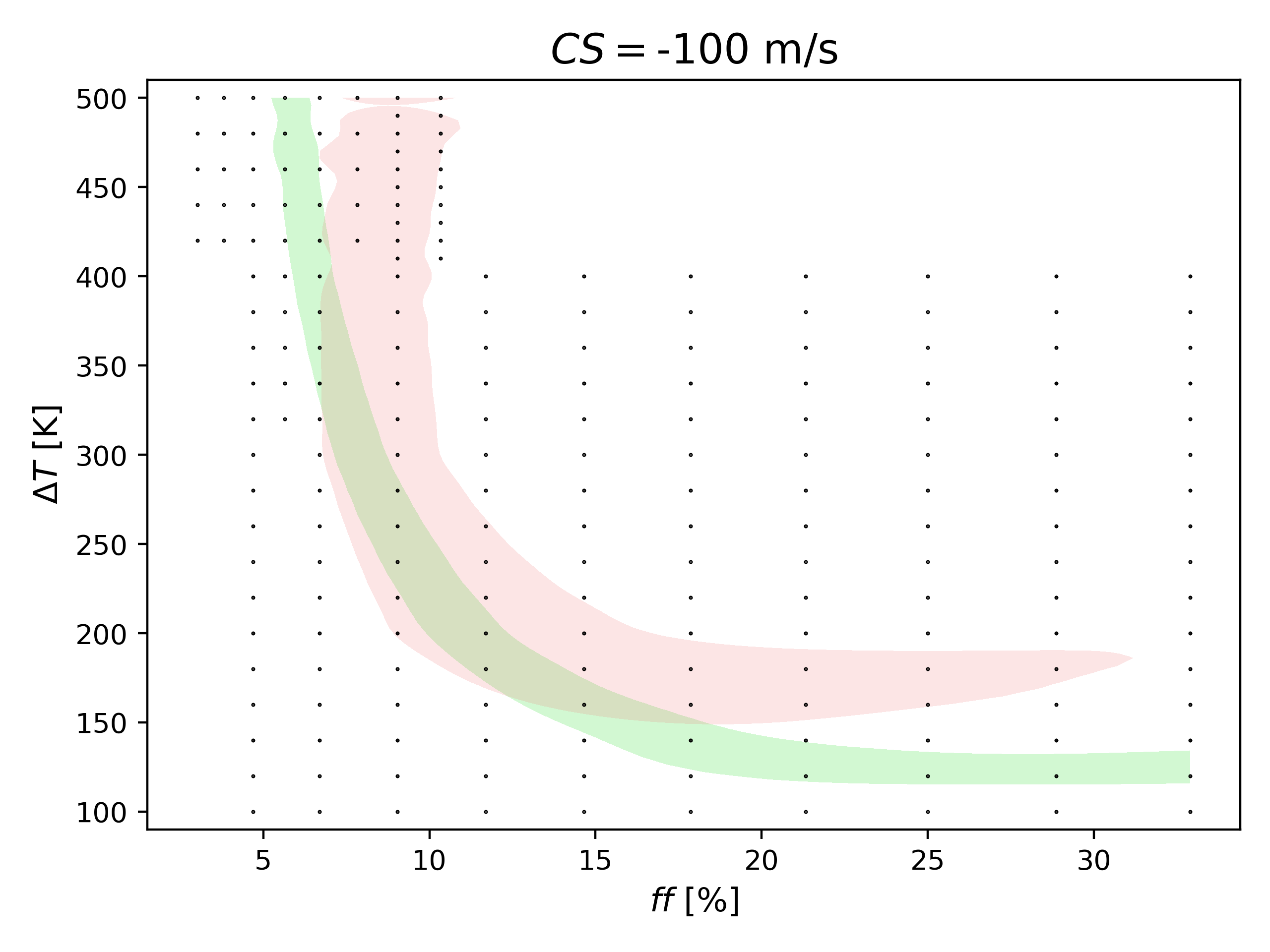}
  \includegraphics[width=0.33\textwidth]{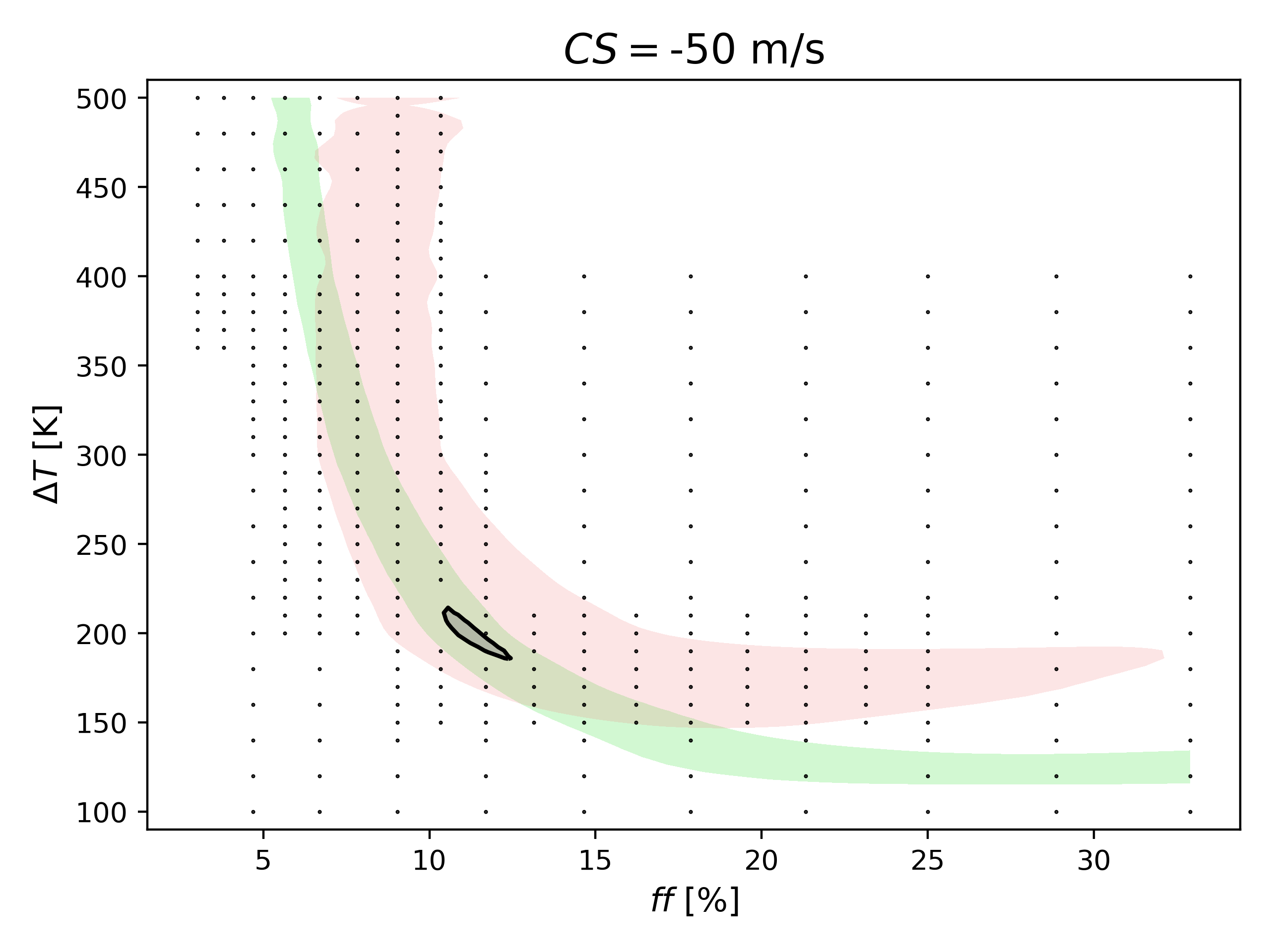}
  \includegraphics[width=0.33\textwidth]{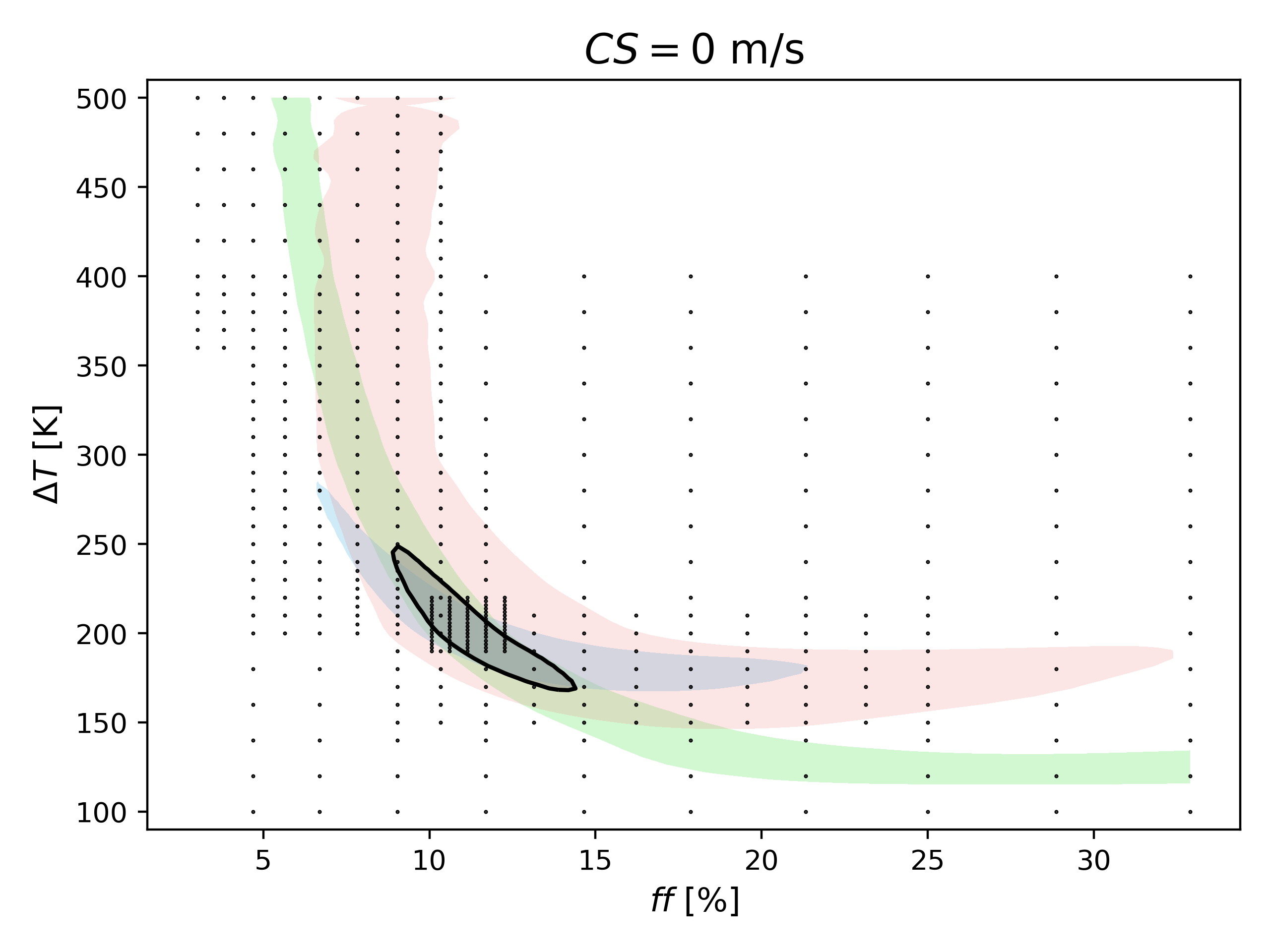}
  \includegraphics[width=0.33\textwidth]{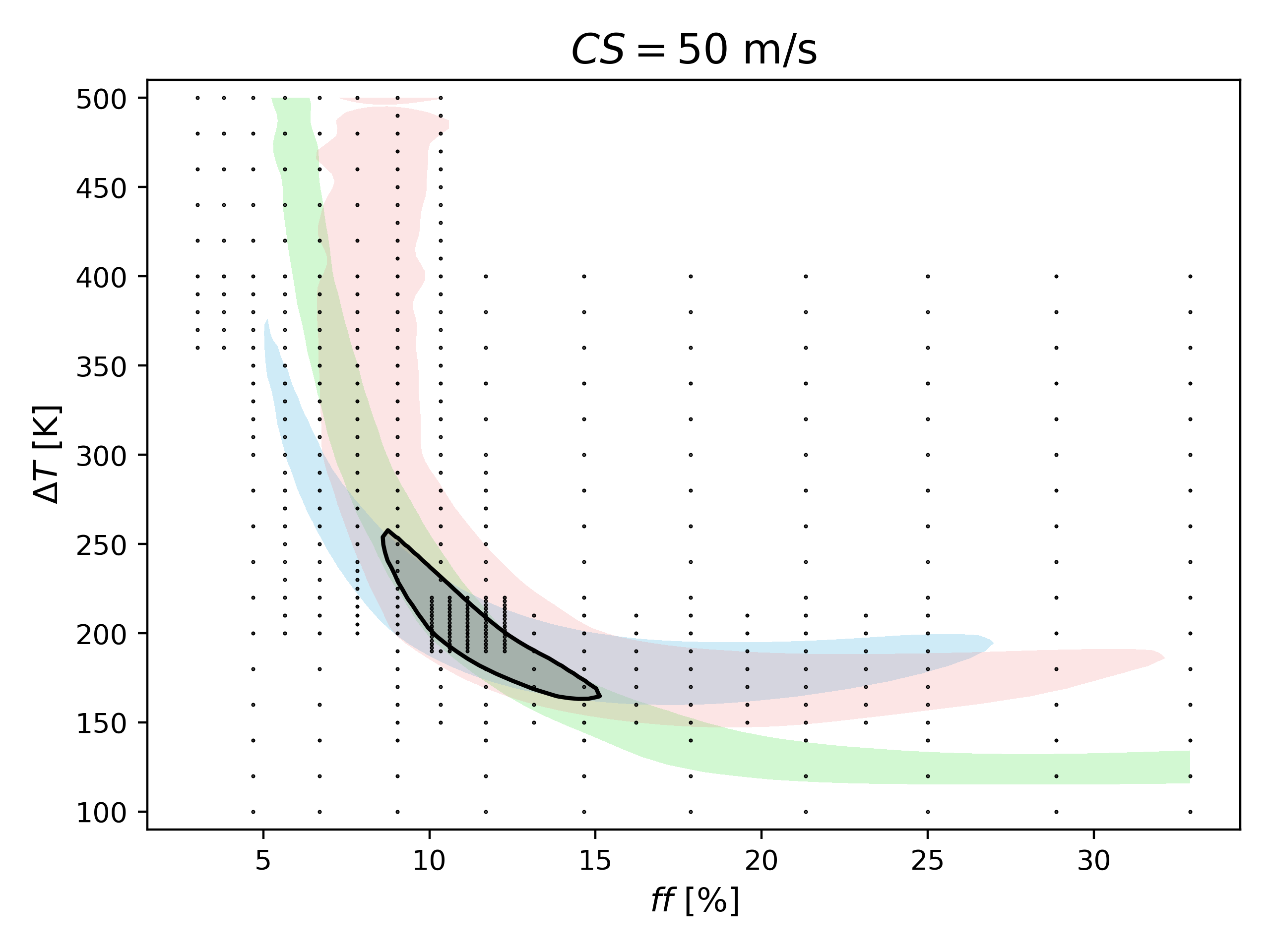}
  \includegraphics[width=0.33\textwidth]{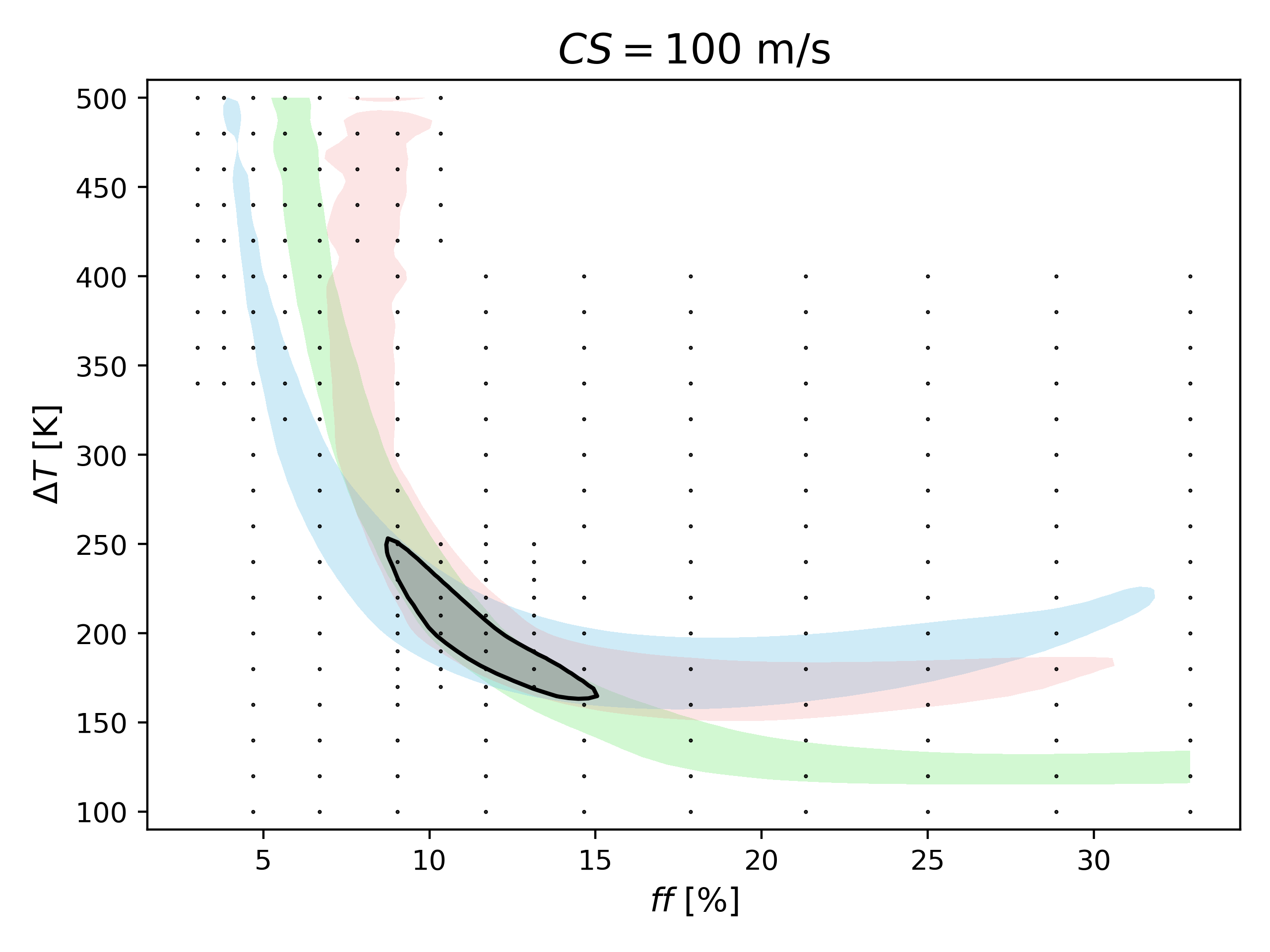}
  \includegraphics[width=0.33\textwidth]{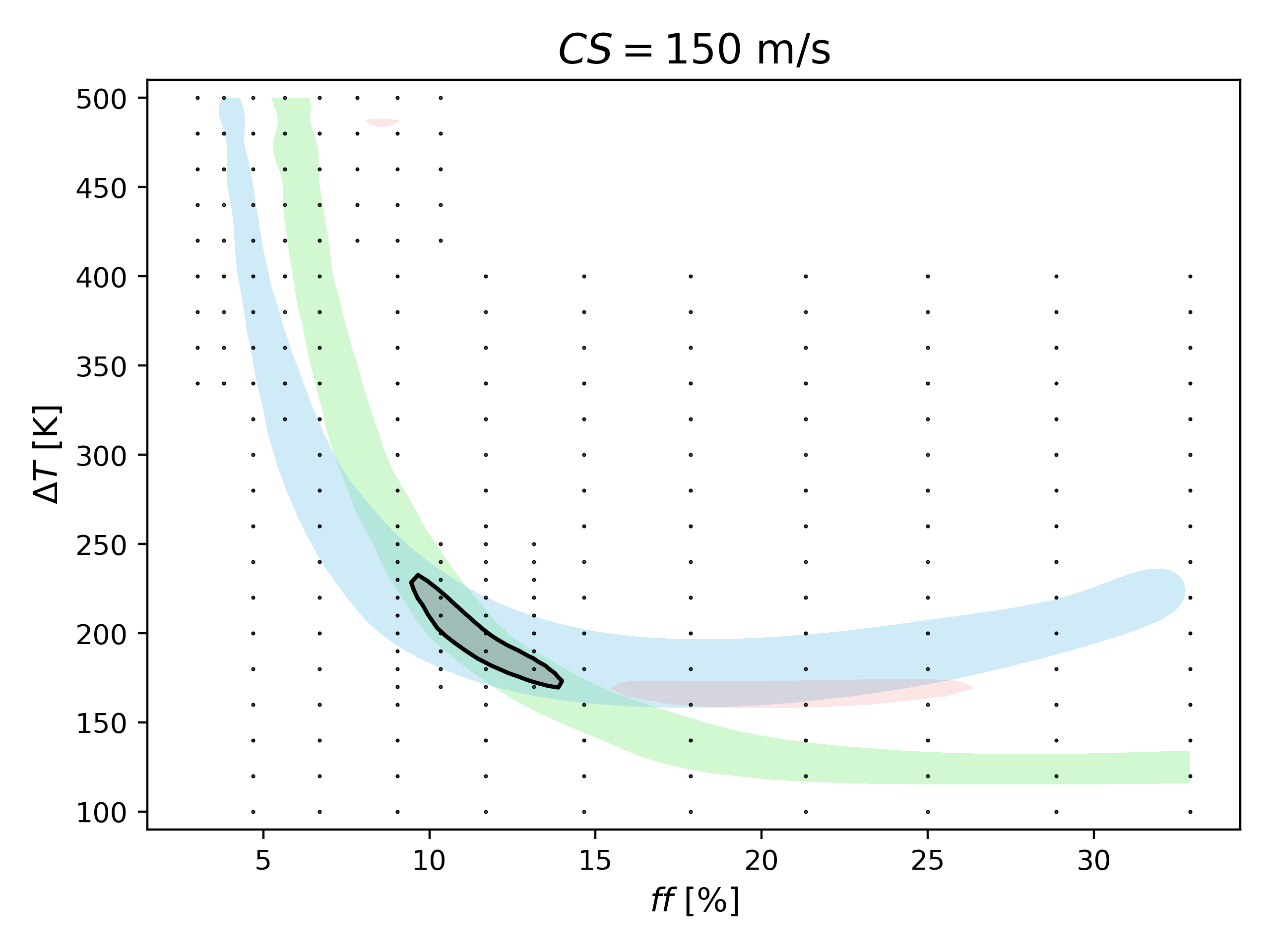}
  \includegraphics[width=0.33\textwidth]{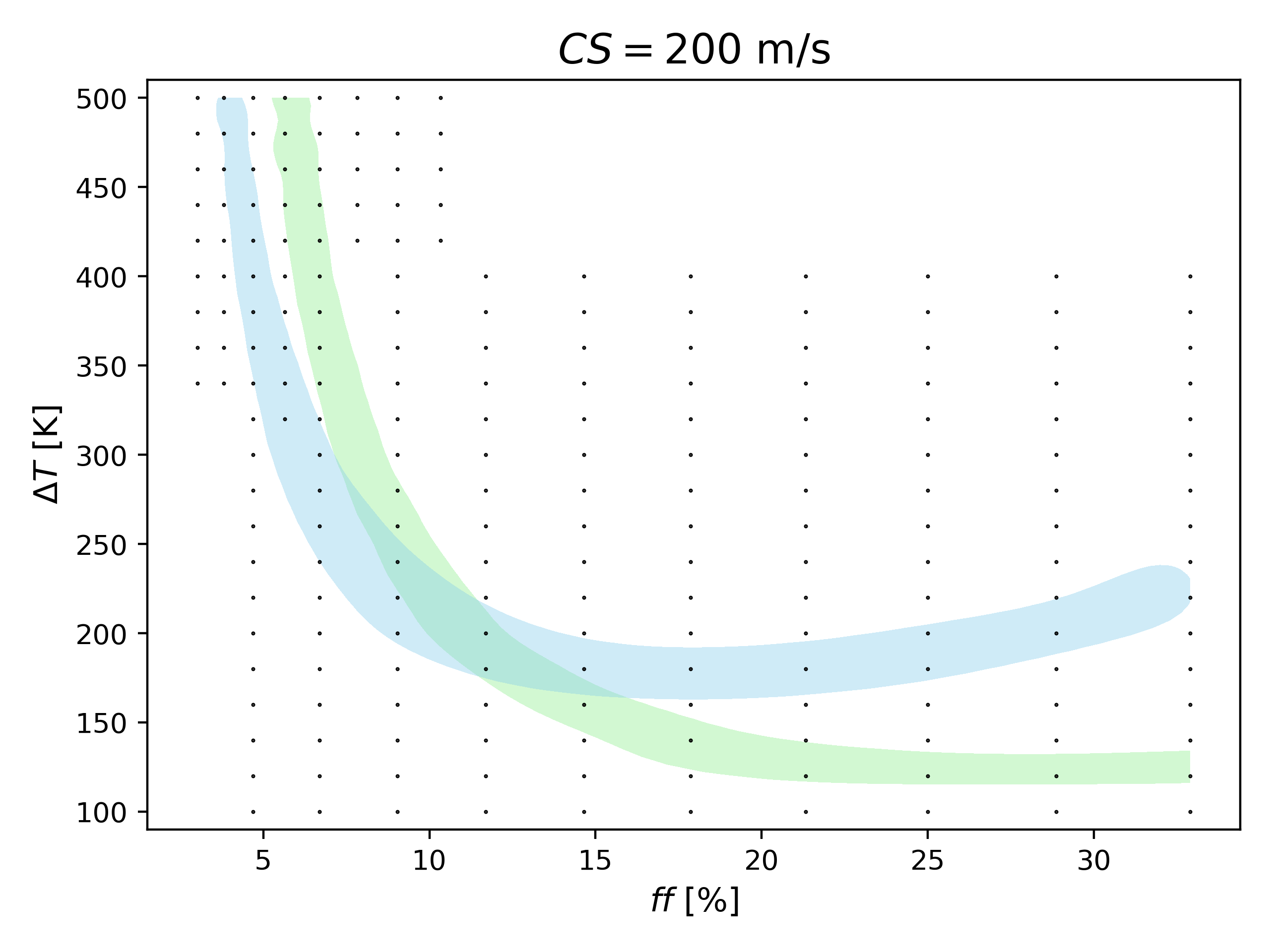}
  \includegraphics[width=0.33\textwidth]{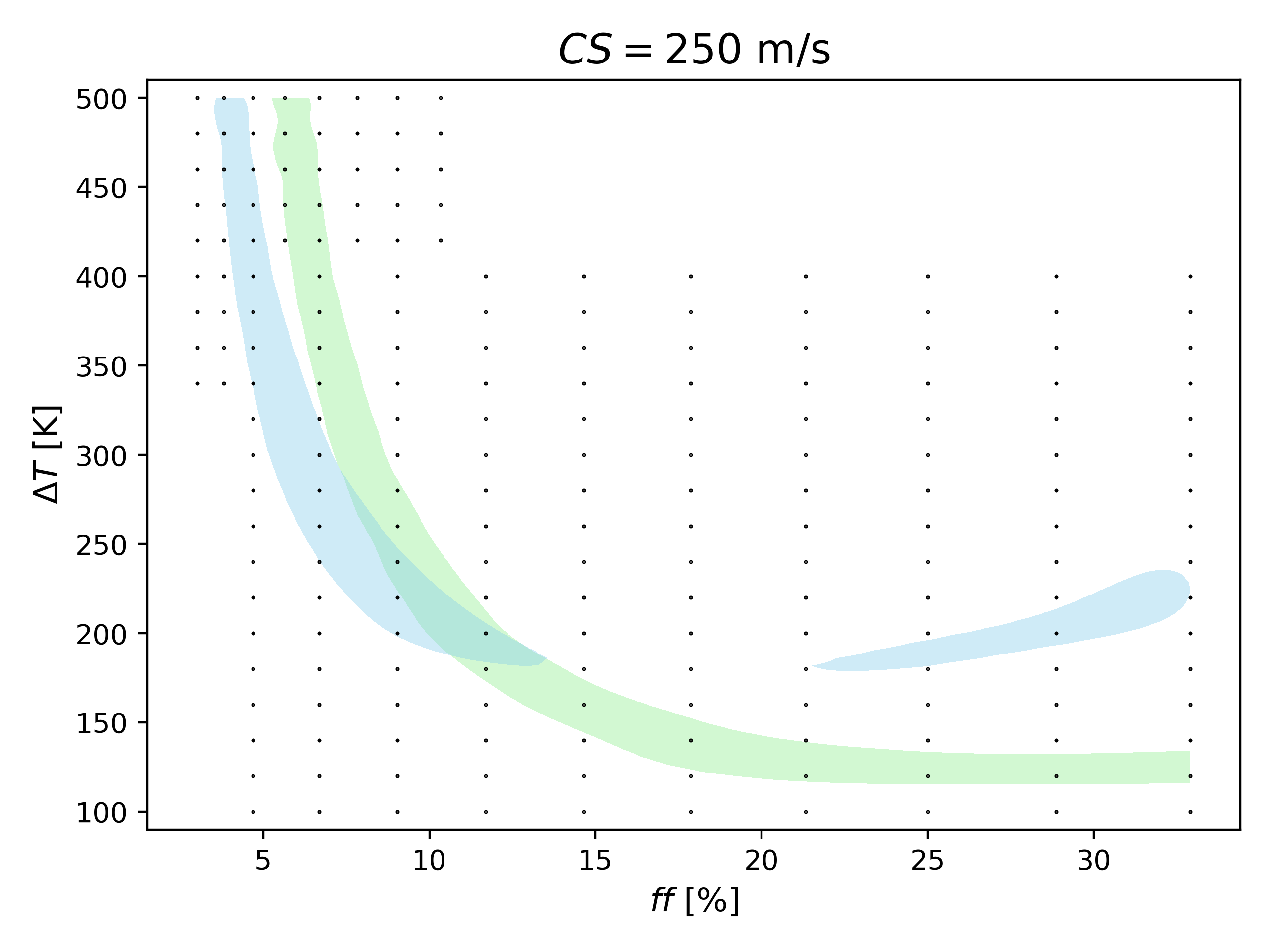}

  \caption{$\Delta \ln L$ contour plots 
  corresponding to the RV (\textit{blue}), CRX (\textit{red}), photometry (\textit{green}), and combined (\textit{black}) fits for different $CS$ slices.  
  The colored regions are within a $\Delta \ln L = 10$ 
  with respect to the best fit among all the $CS$ slices. Black dots indicate the points of contrast temperature and filling factor used to sample the parameter space.}
              \label{slices}
    \end{figure*}

Figure\,\ref{slices} displays the log-likelihood difference with respect to the best model, $\Delta \ln L$, in the $\Delta T$-\textit{ff} plane for different values of $CS$. Dotted symbols correspond to the inspected grid of parameters, which is finer in the regions corresponding to the best fit of observations. The $\Delta \ln L$ corresponding to the RV, CRX, and photometry fits are independently plotted with blue, red, and green contours, respectively, while black contours indicate the joint values. Although photometric fits are independent of the convective shift, they constrain the \textit{ff} and $\Delta T$ correlation very well. On the other hand, both RV and CRX are much more sensitive to $CS$ changes, which breaks the strong \textit{ff}-$\Delta T$ correlation present in the photometry fits.
The different panels in Fig.\,\ref{slices} show that, in spite of the similar solution ranges for $ff$ and $\Delta T$, the CRX provides a stringent constraint on the upper value of $CS$. As a further check, we ran a test excluding the CRX dataset and we found that the uncertainty on the value of $CS$ increases by a factor of $\sim$2. Thus, although the filling factor and the spot temperature contrast can be determined from a simultaneous fit to the RV and photometry only, the CRX provides valuable information to constrain the convective shift. Overlap between photometry, RV, and CRX $\Delta \ln L$ surfaces (i.e., best simultaneous fit to all datasets) occurs around $CS\sim50$\,m\,s$^{-1}$, suggesting a global convective redshift for YZ\,CMi. We note, however, that although the best solutions yield convective redshift, there are acceptable solutions ($\Delta \ln L<10$) located in the convective blueshift region.

\begin{figure}[t]
\centering
\includegraphics[width=\columnwidth]{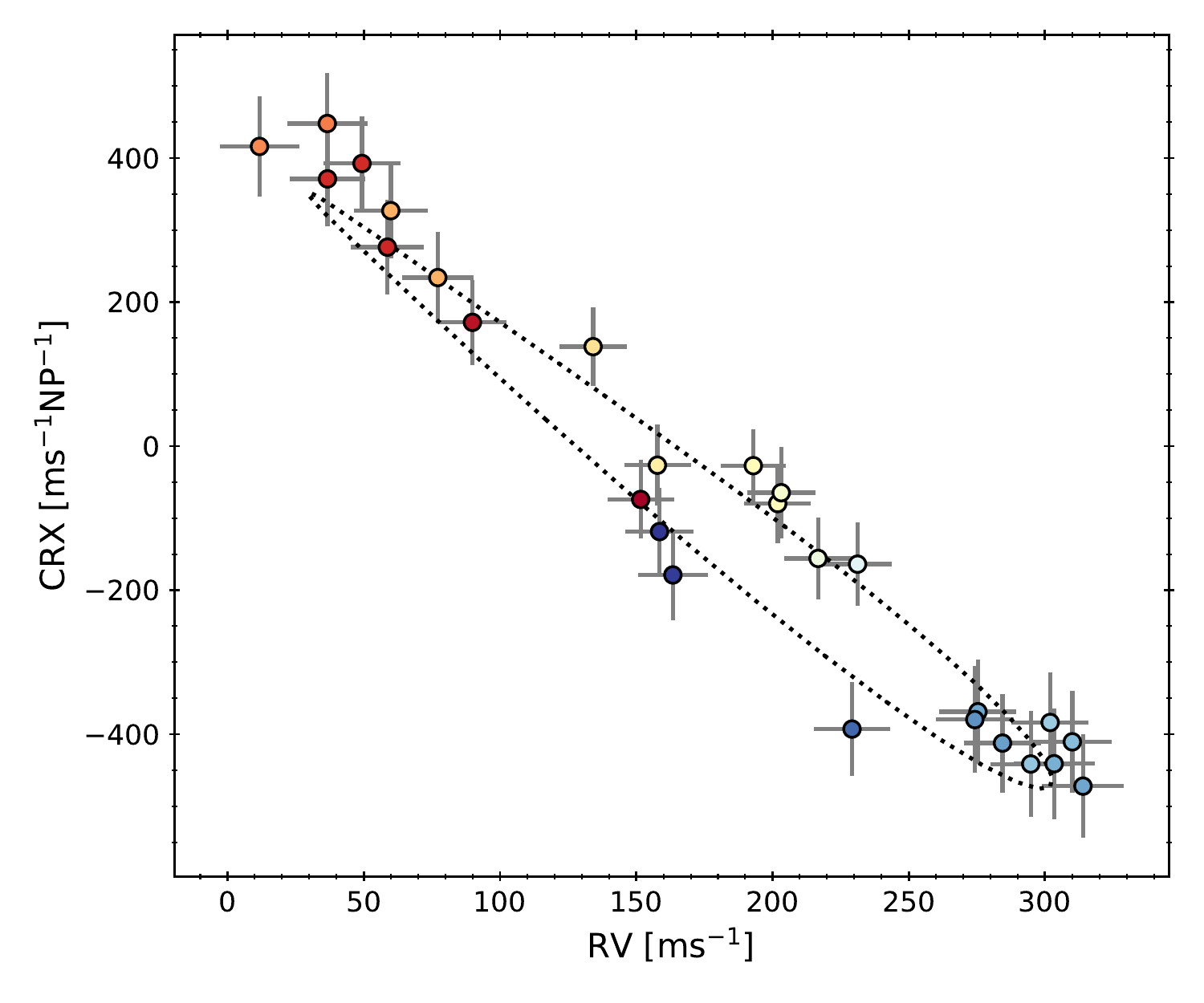}
\caption{Best fit model \textit{(dotted line)} compared to observations \textit{(symbols)} in the RV-CRX parameter space. The colors of the symbols indicate the rotation phase at which each observation was made.}
          \label{corrfit}%
\end{figure}

To estimate the optimal parameters fitting the CARMENES RV and CRX data and the TJO light curve, along with their uncertainties, we interpolated the $\ln L$ hypersurface over the grid used to search for the best solution ($\Delta T$, \textit{ff}, $CS$), also including the adjusted parameters ($\gamma_{\rm RV}$, $\Delta t_{\rm ref}$, $\sigma_{\rm RV}$, $\sigma_{\rm CRX}$, and $\sigma_{\rm phot}$). For this purpose, we used a Gaussian process model based on a squared exponential covariance function, similarly to the likelihood inference explained by \citet{Fleming2018}. Compared to a linear interpolation, this interpolator produces smoother profiles, and can also infer maxima outside the evaluated points. We maximized the likelihood using the Powell method inside the \texttt{scipy.optimize} Python package. Formal uncertainties were derived from the covariance matrix, computed from the Hessian evaluated at the maximum. Table\,\ref{tab:results} lists the parameters of the best fitting model, together with the minimum and maximum values found using other stellar parameters (see below). The RV and CRX fits using these parameters are shown in the top panel of Fig.\,\ref{phasefit}, while the bottom panel shows the fit to the photometry. Figure \ref{corrfit} illustrates the correlation between RV and CRX, showing that the cyclic evolution is well reproduced by the \texttt{StarSim} model.
Comparing Fig.\,\ref{corrfit} with the bottom panels in Fig.\,\ref{example}, YZ\,CMi data show better consistency with the RV-CRX correlation with negative $CS$ than with the $\infty$ shape corresponding to a null $CS$ effect. This correlation can in fact be used as a clear observable of convective blueshift or redshift thanks to the asymmetry introduced toward positive or negative velocities, respectively. Although the bottom-left panel in this figure shows a twisted loop, this also depends on the filling factor and temperature of spots. For YZ\,CMi, the $CS$ is $\sim$6 times smaller than in the simulations in Fig.\,\ref{example}, but the filling factor is also $\sim$6 times larger, causing the untwisted loop.

\begin{table}[t]
\centering
\caption{Best fitting parameters and ranges introduced by the uncertainties on T$_{\rm eff}$ and $i$ to the YZ\,CMi RV, CRX, and photometric datasets.} 
\label{tab:results}
\begin{tabular}{lcccc} 
\hline\hline
\noalign{\smallskip}
\multirow{2}{*}{Parameters} & \multirow{2}{*}{Best model} & \multicolumn{2}{c}{Range} \\
\cmidrule(l){3-4}
 & & Min. & Max. \\
\noalign{\smallskip}
\hline
\noalign{\smallskip}
\textit{ff} [$\%$] & $11.16\pm0.66$  & 9.53 & 13.25  \\
$\Delta T$ [K] & $199.7\pm9.$6  & 178.6 & 273.4  \\
$CS$ [m\,s$^{-1}$] & $56\pm37$ & 7 & 237  \\
$\Delta t_{\rm ref}$ [d] & $1.9879\pm0.0078$ & 1.9735 & 2.0043  \\
$\gamma_{\rm RV}$ [m\,s$^{-1}$] & $ -166.8\pm2.5$ & $-$169.5 & $-$163.6  \\
$\sigma_{\rm RV}$ [m\,s$^{-1}$] & $11.2\pm2.2$ & 8.5 & 17.7 \\
$\sigma_{\rm CRX}$ [m\,s$^{-1}$\,Np$^{-1}$] & $47\pm14$ & 33 & 125 \\
$\sigma_{\rm phot}$ [$\times10^{-3}$] & $6.00\pm0.47$ & 5.45 & 6.67  \\
$\ln L$ & 70.8 & 52.0 & 72.8  \\
\noalign{\smallskip}
\hline
\end{tabular}
\end{table}

The uncertainties of the spot parameters may be underestimated, because we assumed fixed stellar properties. We remind the reader that this is due to the computational effort needed to run the RV simulations, which prevents us from exploring solutions also including stellar properties as free parameters. To study the impact on the parameter uncertainties, we repeated the process to simultaneously fit the RV and CRX data considering a set of values for the effective temperature ($T_{\rm eff}$) of YZ\,CMi and its spin axis inclination ($i$) spanning the reported uncertainties in Table\,\ref{tab:props}. These are the parameters identified to potentially induce changes on the simulated RV. A different $T_{\rm eff}$ could potentially change the temperature and size of spots required to reproduce the observations, while a change in $i$ may also have an impact on the latitude and size of the spots and their imprint on RVs. Table\,\ref{tab:resultsTOT} lists the results of the fits for the set of T$_{\rm eff}$ and $i$ values used for this purpose. All solutions are statistically equivalent ($\Delta \ln L<10$), except for the simulations with an inclination of 52.5\,deg, which start to fail at reproducing the CRX, as indicated by their relatively higher jitter. Additionally, the fit parameters span a wider range than the formal uncertainties previously reported for fixed stellar parameters. We list these additional systematic uncertainties caused by the error bars of the stellar properties in the third column of Table\,\ref{tab:results}.

\begin{table*}[t]
\centering
\caption{Best fit parameters for every $T_{\rm eff}$-$i$ 
set of models.} 
\label{tab:resultsTOT}
\resizebox{\textwidth}{!}{
\begin{tabular}{cccccccccccc} 
\hline\hline
\noalign{\smallskip}
\multicolumn{3}{c}{Model} & \multicolumn{8}{c}{Best fit parameters} & \multirow{3}{*}{$\ln L$}\\
\cmidrule(l){1-3} \cmidrule(l){4-11}
$T_{\rm eff}$ & $i$ & $\lambda_{\rm spot}$ & {\em ff}$_{\rm max}$ 
& $\Delta T$ & $CS$ & $\Delta t_{\rm ref}$ & $\gamma_{\rm RV}$ & $\sigma_{\rm RV}$ & $\sigma_{\rm CRX}$ & $\sigma_{\rm phot}$ & \\
(K) & (deg) & (deg) & (\%) & (K) & (m\,s$^{-1}$) & (d) & (m\,s$^{-1}$) & (m\,s$^{-1}$) & (m\,s$^{-1}$\,Np$^{-1}$) & ($\times 10^{-3}$) \\
\noalign{\smallskip}
\hline
\noalign{\smallskip}
3050 & 21.5 & 73 & 11.66\,$\pm$\,0.78 & 196\,$\pm$\,12 & 49\,$\pm$\,39 & 1.9932\,$\pm$\,0.0080 & $-$167.0\,$\pm$\,2.4 & 10.9\,$\pm$\,2.0 & 50\,$\pm$\,13 & 5.93\,$\pm$\,0.46 & 71.4 \\
3050 & 30.0 & 77 & 11.89\,$\pm$\,0.72 & 188.5\,$\pm$\,9.9 & 57\,$\pm$\,39 & 1.9930\,$\pm$\,0.0080 & $-$167.0\,$\pm$\,2.5 & 11.1\,$\pm$\,2.1 & 49\,$\pm$\,14 & 5.93\,$\pm$\,0.46 & 71.1 \\
3050 & 36.0 & 78 & 11.11\,$\pm$\,0.67 & 194\,$\pm$\,10 & 49\,$\pm$\,39 & 1.9869\,$\pm$\,0.0079 & $-$166.8\,$\pm$\,2.5 & 11.4\,$\pm$\,2.3 & 48\,$\pm$\,14  & 6.00\,$\pm$\,0.47 & 69.9 \\
3050 & 45.0 & 79 & 10.51\,$\pm$\,0.71 & 210\,$\pm$\,13 & 62\,$\pm$\,24 & 1.9839\,$\pm$\,0.0079 & $-$166.8\,$\pm$\,2.6 & 12.0\,$\pm$\,2.3 & 60\,$\pm$\,14 & 6.09\,$\pm$\,0.48 & 63.8 \\
3050 & 52.5 & 80 & 9.96\,$\pm$\,0.43 & 249.7\,$\pm$\,8.8 & 68\,$\pm$\,61 & 1.9821\,$\pm$\,0.0086 & $-$166.6\,$\pm$\,3.0 & 14.4\,$\pm$\,3.3 & 84\,$\pm$\,20 & 6.17\,$\pm$\,0.50 & 52.0 \\
3100 & 21.5 & 73 & 11.65\,$\pm$\,0.48 & 202.3\,$\pm$\,7.2 & 56\,$\pm$\,37 & 1.9941\,$\pm$\,0.0078 & $-$167.0\,$\pm$\,2.4 & 10.9\,$\pm$\,2.0 & 50\,$\pm$\,13 & 5.93\,$\pm$\,0.46 & 71.6  \\
3100 & 30.0 & 77 & 12.33\,$\pm$\,0.92 & 190.5\,$\pm$\,9.6 & 70\,$\pm$\,37 & 1.9931\,$\pm$\,0.0081 & $-$167.1\,$\pm$\,2.4 & 10.7\,$\pm$\,2.0 & 49\,$\pm$\,13 & 5.93\,$\pm$\,0.46 & 71.9\\
3100\tablefootmark{a} & 36.0\tablefootmark{a} & 78\tablefootmark{a} & 11.16\,$\pm$\,0.66 & 199.7\,$\pm$\,9.6 & 56\,$\pm$\,37 & 1.9879\,$\pm$\,0.0078 & $-$166.8\,$\pm$\,2.5 & 11.2\,$\pm$\,2.2 & 47\,$\pm$\,14 & 6.00\,$\pm$\,0.47 & 70.8 \\
3100 & 45.0 & 79 & 10.68\,$\pm$\,0.63 & 214.7\,$\pm$\,9.4 & 79\,$\pm$\,22 & 1.9855\,$\pm$\,0.0080 & $-$166.8\,$\pm$\,2.5 & 11.5\,$\pm$\,2.3 & 61\,$\pm$\,15 & 6.06\,$\pm$\,0.48 & 64.8 \\
3100 & 52.5 & 80 & 10.26\,$\pm$\,0.67 & 257\,$\pm$\,14 & 167\,$\pm$\,61 & 1.9869\,$\pm$\,0.0089 & $-$166.9\,$\pm$\,2.5 & 11.6\,$\pm$\,2.2 & 103\,$\pm$\,22 & 6.09\,$\pm$\,0.49 & 53.3 \\
3150 & 21.5 & 73 & 11.51\,$\pm$\,0.90 & 212\,$\pm$\,14 & 74\,$\pm$\,39 & 1.9962\,$\pm$\,0.0081 & $-$167.0\,$\pm$\,2.4 & 10.7\,$\pm$\,1.9 & 55\,$\pm$\,13 & 5.91\,$\pm$\,0.46 & 70.7\\
3150 & 30.0 & 77 & 12.06\,$\pm$\,0.74 & 201.0\,$\pm$\,9.8 & 86\,$\pm$\,35 & 1.9947\,$\pm$\,0.0079 & $-$167.0\,$\pm$\,2.4 & 10.6\,$\pm$\,1.9 & 49\,$\pm$\,13 & 5.92\,$\pm$\,0.46 & 72.8\\
3150 & 36.0 & 78 & 11.76\,$\pm$\,0.81 & 200\,$\pm$\,11 & 93\,$\pm$\,32 & 1.9930\,$\pm$\,0.0078 & $-$167.0\,$\pm$\,2.3 & 10.4\,$\pm$\,1.9 & 49\,$\pm$\,13 & 5.95\,$\pm$\,0.46 & 72.4\\
3150 & 45.0 & 79 & 11.01\,$\pm$\,0.86 & 215\,$\pm$\,13 & 132\,$\pm$\,39 & 1.9932\,$\pm$\,0.0083 & $-$166.9\,$\pm$\,2.3 & 10.4\,$\pm$\,2.0 & 64\,$\pm$\,15 & 5.98\,$\pm$\,0.47 & 67.1 \\
3150 & 52.5 & 80 & 10.55\,$\pm$\,0.46 & 253.0\,$\pm$\,9.5 & 180\,$\pm$\,57 & 1.9881\,$\pm$\,0.0082 & $-$167.0\,$\pm$\,2.5 & 11.3\,$\pm$\,2.4 & 91\,$\pm$\,20 & 6.11\,$\pm$\,0.49 & 56.3 \\
\hline
\end{tabular}
}
\tablefoot{
\tablefoottext{a}{Stellar parameters used as the best model in Table \ref{tab:results}}
}
\end{table*}

\section{Discussion and conclusions}\label{sec:discussion}

Our model for YZ\,CMi indicates that this active star has at least a prominent spot with a filling factor in the $\sim$9.5--13.3\,\% range. The value that we found is compatible with previous values in the literature by \cite{Zboril2003} and \cite{Alekseev2017}, who suggested filling factors of 10--25\,\% and 10--38\,\%, respectively. The significantly broader (and less precise) ranges reported in these works most likely arise from the use of only light curves, which, as we showed in Fig.\,\ref{slices}, produce a strong degeneracy in the \textit{ff}-$\Delta T$ plane. However, because of the different adopted stellar inclinations, contrast temperatures, and spot models used in these two works, caution should be taken when
performing comparisons. They used stellar inclinations of 60--70\,deg, \cite{Zboril2003} fixed the contrast temperature to 500\,K, and \cite{Alekseev2017} used two spot belts. Furthermore, the spot size and latitude that we estimate are consistent with a spot covering a large fraction of the visible pole, in agreement with the results obtained by Zeeman-Doppler imaging \citep{Morin2008}.

From our analysis, we constrain the temperature difference between the photosphere and the spot to a $1\sigma$ confidence range of $\sim$179--273\,K. These values are in close agreement with the $\Delta T$ results by \cite{Alekseev2017}, and consistent with the commonly used empirical calibration by \cite{Berdyugina2005} and \cite{Andersen2015}. However, our results are better constrained due to the simultaneous fit of photometric, RV, and CRX time series covering several wavelength bands that allow the breaking of the spot temperature and the filling factor degeneracy. Furthermore, the scarcity of spot crossing events on the $\sim$4000 M dwarfs in the {\em Kepler} catalog could be indicative of a low contrast ratio for the spots on the photospheres of these stars (i.e., smaller $\Delta T$), assuming a heterogeneous distribution of spots \citep{Andersen2015}. The study of multiband photometric observations also points toward a smaller temperature difference for late-type stars than for solar-like stars \citep{Mallon2018}. 

\begin{figure}[t]
\centering
\includegraphics[width=\columnwidth]{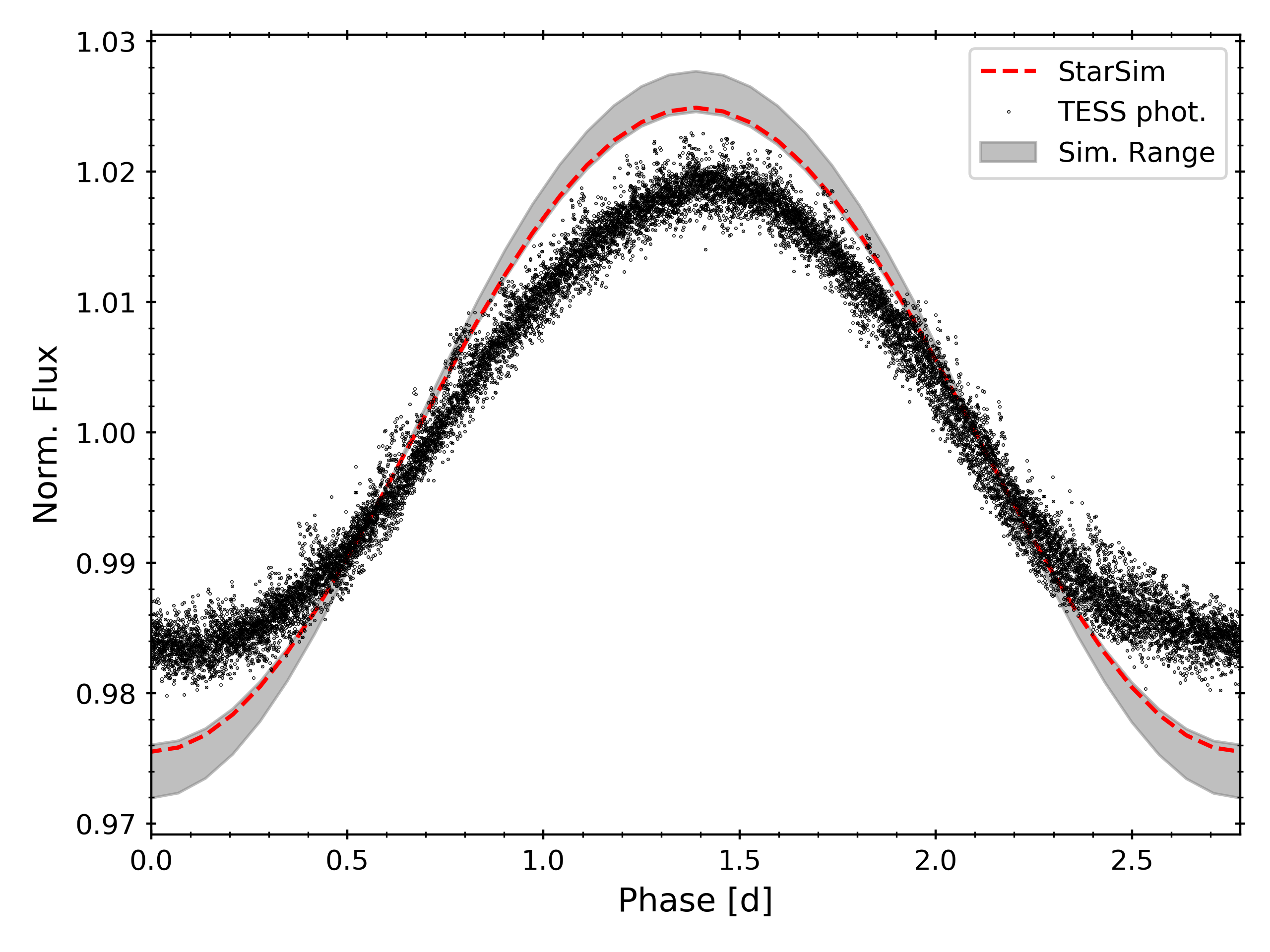}
\caption{Phase-folded {\em TESS} photometric observations ({\em black dots} compared to the \texttt{StarSim} simulation ({\em dashed line}) using the best model values and the {\em TESS}-band filter. The gray shaded region corresponds to the range defined by all the best model simulations in Table \ref{tab:resultsTOT}.}
          \label{TESS}%
\end{figure}

We note that differences in temperature and filling factors available in the literature may also be due to the evolution of stellar activity features. For instance, as a further check, we compared our simulations with the two-minute cadence photometry available for YZ\,CMi from the {\em Transiting Exoplanet Survey Satellite} \cite[{\em TESS},][]{Ricker2015}, obtained between January and February 2019, two years after our CARMENES dataset. Figure\,\ref{TESS} shows this light curve along with the model corresponding to the best fitting parameters listed in Table \ref{tab:results}, and with the range defined by all the models in Table \ref{tab:resultsTOT}. Although the overall aspect is similar, the {\em TESS} light curve has a lower amplitude compared to our models. This difference can be reproduced by reducing the filling factor by $\sim3.5$\,\% or with a temperature contrast 40\,K lower, and may be due to the time evolution of spots. Interestingly, the phase and shape of the modulation is consistent with our model, which may point to a long-lived active region that does not change its position on the stellar surface significantly.

Finally, the simultaneous fit to RV, CRX, and photometric time series results in a global convective shift for YZ\,CMi in the range between +7 and +237\,m\,s$^{-1}$, in contrast with the value of $-$300\,m\,s$^{-1}$ estimated for the Sun. Interestingly, this result means that the convection effect produces a net redshift, not a blueshift, although this result must be taken with caution given the large uncertainty. Such a possibility was already suggested by \cite{Kurster2003} based on the anti-correlation between the H$\alpha$ line strength and the RV for Barnard's Star, indicating an increase in the blueshift when the coverage of the star with plage regions increases. Similar conclusions are suggested by 3D hydrodynamic simulations, which also predict a very small net convective blueshift for late K-type stars, but they increase with the effective temperature of the star until they reach a blueshift of 300\,m\,s$^{-1}$ for F-type dwarfs \citep{Allende2013}. \cite{Meunier2017} described a similar trend from the estimation of the convective blueshift using Fe and Ti lines of 360 F7--K4 stars. For the particular case of M-dwarf stars, magnetohydrodynamic simulations suggest that convective motions are less vigorous and that the average granule velocity shift is also smaller \citep{Beeck2013a,Beeck2013b}.

To conclude, the results of our analysis reveal that a simultaneous fit to light and RV curves for several wavelength bands represents a novel approach to estimating not only properties of stellar activity, such as the spot filling factor and temperature difference (breaking their degeneracy), but also the shift of radial velocities due to convective motions. From the particular case of YZ\,CMi presented here, and its comparison to the Sun, we conclude that the absolute convective shift may be reversed toward redshift for M-dwarf stars. Finally, the chromatic index is not affected by the Keplerian orbital motion of exoplanets seen in the radial velocity, and can therefore provide vital information to disentangle exoplanet signals from stellar activity effects.

\begin{acknowledgements}
Based on observations collected at the Centro Astron\'omico Hispano Alem\'an (CAHA) at Calar Alto, operated jointly by the Junta de Andaluc\'ia and the Instituto de Astrof\'isica de Andaluc\'ia (CSIC). CARMENES is funded by the German Max-Planck-Gesellschaft (MPG), the Spanish Consejo Superior de Investigaciones Cient\'ificas (CSIC), the European Union through FEDER/ERF FICTS-2011-02 funds, and the members of the CARMENES Consortium (Max-Planck-Institut f\"ur Astronomie, Instituto de Astrof\'isica de Andaluc\'ia, Landessternwarte K\"onigstuhl, Institut de Ci\`encies de l'Espai, Insitut f\"ur Astrophysik G\"ottingen, Universidad Complutense de Madrid, Th\"uringer Landessternwarte Tautenburg, Instituto de Astro\'isica de Canarias, Hamburger Sternwarte, Centro de Astrobiolog\'ia and Centro Astron\'omico Hispano-Alem\'an), with additional contributions by the Spanish Ministry of Economy, the German Science Foundation through the Major Research Instrumentation Programme and DFG Research Unit FOR2544 ``Blue Planets around Red Stars'', the Klaus Tschira Stiftung, the states of Baden-W\"urttemberg and Niedersachsen, and by the Junta de Andaluc\'ia. 
We acknowledge financial support from the Spanish Agencia Estatal de Investigaci\'on of the Ministerio de Ciencia e Innovaci\'on and the European FEDER/ERF funds through projects 
  AYA2016-79425-C3-1/2/3-P,
  PGC2018-098153-B-C33, 
  BES-2017-080769
and the Centre of Excellence ``Severo Ochoa'' and ``Mar\'ia de Maeztu'' awards to the Instituto de Astrof\'isica de Canarias (SEV-2015-0548), Instituto de Astrof\'isica de Andaluc\'ia (SEV-2017-0709), and Centro de Astrobiolog\'ia (MDM-2017-0737),
the Secretaria d'Universitats i Recerca del Departament d'Empresa i Coneixement de la Generalitat de Catalunya and the Ag\`encia de Gesti\'o d'Ajuts Universitaris i de Recerca of the Generalitat de Catalunya, with additional funding from the European FEDER/ERF funds, \emph{L'FSE inverteix en el teu futur}, the Generalitat de Catalunya/CERCA programme, and from NASA Grant NNX17AG24G.
This work makes use of data from the 80\,cm Telescopi Joan Or\'o (TJO) of the Montsec Astronomical Observatory (OAdM), owned by the Generalitat de Catalunya and operated by the Institut d'Estudis Espacials de Catalunya (IEEC), and includes data collected by the {\em TESS} mission. Funding for the {\em TESS} mission is provided by the NASA Explorer Program. 

\end{acknowledgements}

\bibliographystyle{aa} 
\bibliography{bibtex.bib}

\begin{appendix}
\section{Long Tables}

\begin{table}[ht]
\centering
\caption{Date of observation, radial velocities, and chromatic index (with their associated errors) of the YZ\,CMi CARMENES spectroscopic observations.} 
\label{tab:data}
\begin{tabular}{ccc} 
\hline\hline
\noalign{\smallskip}
HJD -- 2450000 [d] & RV [m\,s$^{-1}$] & CRX [m\,s$^{-1}$\,Np$^{-1}$]\\
\noalign{\smallskip}
\hline
\noalign{\smallskip}
7655.71 & $163.5\pm4.6$ & $-179\pm41$ \\ 
7673.66 & $157.8\pm2.8$ & $-27\pm29$ \\ 
7689.73 & $36.6\pm8.7$ & $448\pm51$ \\ 
7692.72 & $77.2\pm5.3$ & $234\pm41$ \\ 
7699.65 & $275.4\pm7.6$ & $-369\pm54$ \\ 
7704.60 & $231.2\pm3.9$ & $-164\pm32$ \\ 
7735.61 & $310.0\pm8.1$ & $-411\pm53$ \\ 
7755.69 & $158.6\pm3.8$ & $-119\pm37$ \\ 
7756.56 & $59.9\pm6.6$ & $327\pm45$ \\ 
7760.50 & $301.9\pm7.6$ & $-384\pm51$ \\ 
7761.41 & $89.9\pm4.1$ & $172\pm35$ \\ 
7762.62 & $201.9\pm2.8$ & $-80\pm26$ \\ 
7763.53 & $284.4\pm7.9$ & $-412\pm49$ \\ 
7779.46 & $216.7\pm3.7$ & $-156\pm30$ \\ 
7786.49 & $58.7\pm6.1$ & $276\pm45$ \\ 
7788.48 & $314.0\pm9.1$ & $-472\pm54$ \\ 
7790.49 & $203.2\pm4.1$ & $-65\pm41$ \\ 
7798.48 & $134.2\pm3.3$ & $138\pm26$ \\ 
7800.41 & $49.4\pm7.7$ & $392\pm45$ \\ 
7815.36 & $193.0\pm1.6$ & $-27\pm16$ \\ 
7822.36 & $151.6\pm2.9$ & $-74\pm26$ \\ 
7830.42 & $229.2\pm7.4$ & $-393\pm44$ \\ 
7849.36 & $294.8\pm8.6$ & $-441\pm56$ \\ 
7852.42 & $274.3\pm7.7$ & $-380\pm57$ \\ 
7856.34 & $11.7\pm8.4$ & $416\pm50$ \\ 
7863.38 & $303.4\pm8.9$ & $-441\pm60$ \\ 
7875.36 & $36.7\pm7.1$ & $371\pm44$ \\ 
\hline
\end{tabular}

\end{table}

\begin{table}[ht]
\centering
\caption{Date of observation and normalised flux (with associated errors) of the YZ\,CMi TJO photometric observations.} 
\label{tab:dataph}
\begin{tabular}{cc} 
\hline\hline
\noalign{\smallskip}
HJD -- 2450000 [d] & Norm. flux \\ 
\noalign{\smallskip}
\hline
\noalign{\smallskip}
7688.643 & $0.9808\pm0.0005$\\
7689.639 & $0.9827\pm0.0005$\\
7692.617 & $1.0003\pm0.0006$\\
7693.567 & $1.0348\pm0.0005$\\
7693.651 & $1.0383\pm0.0005$\\
7694.569 & $0.9655\pm0.0007$\\
7694.654 & $0.9539\pm0.0005$\\
7695.572 & $1.0207\pm0.0009$\\
7695.573 & $1.0224\pm0.0011$\\
7696.561 & $1.0122\pm0.0006$\\
7696.646 & $0.9997\pm0.0006$\\
7699.646 & $0.9854\pm0.0005$\\
7700.567 & $0.9615\pm0.0005$\\
7702.543 & $0.9833\pm0.0005$\\
7708.527 & $0.9572\pm0.0007$\\
7708.609 & $0.9489\pm0.0004$\\
7723.526 & $1.0393\pm0.0005$\\
7723.624 & $1.0371\pm0.0005$\\
7724.593 & $0.9967\pm0.0011$\\
7724.594 & $0.9930\pm0.0005$\\
7725.570 & $0.9730\pm0.0010$\\
7729.620 & $1.0390\pm0.0012$\\
7730.468 & $0.9833\pm0.0006$\\
7730.469 & $0.9769\pm0.0012$\\
7731.467 & $1.0149\pm0.0006$\\
7734.495 & $1.0202\pm0.0007$\\
7734.562 & $1.0188\pm0.0007$\\
7734.577 & $1.0327\pm0.0008$\\
7735.456 & $1.0169\pm0.0008$\\
7736.536 & $0.9608\pm0.0009$\\
7741.445 & $0.9859\pm0.0005$\\
7741.536 & $0.9658\pm0.0005$\\
7742.454 & $0.9954\pm0.0005$\\
7742.540 & $0.9913\pm0.0005$\\
7746.518 & $1.0185\pm0.0005$\\
7746.618 & $1.0133\pm0.0007$\\
7748.474 & $1.0318\pm0.0005$\\
7749.447 & $1.0068\pm0.0006$\\
7749.448 & $1.0096\pm0.0013$\\
7750.533 & $0.9696\pm0.0005$\\
7751.530 & $1.0450\pm0.0006$\\
7751.531 & $1.0458\pm0.0011$\\
7752.509 & $0.9830\pm0.0006$\\
7752.510 & $0.9827\pm0.0007$\\
7752.606 & $0.9746\pm0.0005$\\
7753.471 & $0.9873\pm0.0008$\\
7756.641 & $1.0318\pm0.0005$\\
7758.461 & $0.9522\pm0.0004$\\
7758.580 & $0.9557\pm0.0006$\\
7759.423 & $1.0204\pm0.0005$\\
7759.508 & $1.0228\pm0.0005$\\
7759.514 & $1.0246\pm0.0005$\\
7760.396 & $1.0224\pm0.0005$\\
7760.433 & $1.0126\pm0.0005$\\
\hline
\end{tabular}
\end{table}

\addtocounter{table}{-1}
\begin{table}[ht]
\centering
\caption{Continued.}
\begin{tabular}{cc} 
\hline\hline
\noalign{\smallskip}
HJD -- 2450000 [d] & Norm. flux \\ 
\noalign{\smallskip}
\hline
\noalign{\smallskip}
7760.523 & $1.0015\pm0.0009$\\
7761.394 & $0.9641\pm0.0005$\\
7761.433 & $0.9624\pm0.0011$\\
7762.382 & $1.0414\pm0.0006$\\
7762.576 & $1.0411\pm0.0005$\\
7763.386 & $0.9999\pm0.0023$\\
7764.463 & $0.9722\pm0.0006$\\
7768.459 & $1.0368\pm0.0006$\\
7768.518 & $1.0264\pm0.0006$\\
7768.544 & $1.0301\pm0.0006$\\
7795.429 & $1.0149\pm0.0015$\\
7804.485 & $1.0538\pm0.0005$\\
7805.487 & $0.9620\pm0.0005$\\
7807.390 & $1.0214\pm0.0006$\\
7807.523 & $1.0124\pm0.0007$\\
7809.450 & $1.0215\pm0.0005$\\
7809.541 & $1.0282\pm0.0006$\\
7811.510 & $0.9641\pm0.0012$\\
7813.405 & $0.9986\pm0.0006$\\
7813.406 & $0.9975\pm0.0008$\\
7819.468 & $0.9528\pm0.0009$\\
7819.469 & $0.9525\pm0.0007$\\
7821.367 & $1.0319\pm0.0018$\\
7821.472 & $1.0143\pm0.0022$\\
7822.389 & $0.9565\pm0.0007$\\
7822.474 & $0.9583\pm0.0010$\\
7827.405 & $0.9807\pm0.0010$\\
7827.406 & $0.9829\pm0.0008$\\
7830.419 & $0.9633\pm0.0011$\\
7830.495 & $0.9717\pm0.0006$\\
7831.419 & $1.0008\pm0.0007$\\
7834.442 & $1.0279\pm0.0011$\\
7836.439 & $0.9720\pm0.0005$\\
7843.422 & $1.0434\pm0.0010$\\
7857.405 & $1.0412\pm0.0006$\\
\hline
\end{tabular}
\end{table}

\end{appendix}

\end{document}